\documentclass[]{aa}
\usepackage{graphicx,natbib}
\usepackage{lipsum}
\usepackage{latexsym,amsmath,tabularx,amssymb}
\usepackage{tablefootnote}
\usepackage{mathrsfs}
\usepackage[latin1]{inputenc}
\usepackage[T1]{fontenc}
\bibpunct{(}{)}{;}{a}{}{,}
\usepackage[varg]{txfonts}
\include{def}

\usepackage{bm}
\usepackage{units}
\usepackage{color}
\usepackage{url}

\newcommand{\ME}{M$_{\oplus}$} 
\newcommand{\RE}{R_{\oplus}} 
\newcommand{\Mearth}{\text{M}_{\oplus}} 
\newcommand{\Msun}{M_{\odot} }
\newcommand{\rhocgs}{g/cm$^3$}
\newcommand{\gcmcuad}{g/cm$^2$}

\newcommand{\Mscat}{M_{\text{scat}}}
\newcommand{\tscat}{t_{\text{scat}}}

\newcommand{\Mcore}{M_{\text{core}}}

\newcommand{\MHHe}{M_{\text{HHe}}}

\newcommand{\Mz}{M_{\text{Z}}}

\newcommand{\Mzdot}{\dot{M}_{\text{Z}}}
\newcommand{\MzdotNG}{\dot{M}_{\text{Z, no gap}}}
\newcommand{\MzdotG}{\dot{M}_{\text{Z, gap}}}

\newcommand{\Mdothydro}{\dot{M}_{\rm gas, disk}}

\newcommand{\Rp}{R_{\text{P}}}
\newcommand{\Rh}{R_{\rm H}}
\newcommand{\Mp}{M_{\text{P}}}
\newcommand{\Miso}{M_{\text{iso}}}

\newcommand{\tcross}{t_{\text{cross}}}

\newcommand{\MdotHHe}{\dot{M}_{\text{HHe}}}
\newcommand{\Mcross}{M_{\text{cross}}}

\newcommand{\rp}{r_{\text{p}}}
\newcommand{\rhop}{\rho_{\text{p}}}

\newcommand{\SigmaG}{\Sigma_{\rm gas}}

\begin{document}

\title{Jupiter's heavy-element enrichment expected from formation models}



 \titlerunning{Origin of Jupiter's heavy elements}
\authorrunning{Venturini \& Helled}

\author{Julia Venturini \inst{1}, Ravit Helled \inst{2}}
\offprints{J. Venturini}
\institute{International Space Science Institute, Hallerstrasse 6, CH-3012 , Bern, Switzerland, 
\and
{Institute for Computational Science, University of Zurich, Winterthurerstrasse 190, CH-8057, Zurich, Switzerland} \\
        \email{julia.venturini@issibern.ch}}

\abstract
{}
{The goal of this work is to investigate Jupiter's growth focusing on the amount of heavy elements accreted by the planet, and its comparison with recent structure models of Jupiter.}
{Our model assumes an initial core growth dominated by pebble accretion, 
and a second growth phase that is characterized by a moderate accretion of both planetesimals and gas. The third phase is dominated by runaway gas accretion during which the planet becomes detached from the disk. 
The second and third phases are computed in detail, considering two different prescriptions for the planetesimal accretion and fits from hydrodynamical studies to compute the gas accretion in the detached phase.}
{In order for Jupiter to consist of $\sim$20-40  $\Mearth$ of heavy elements as suggested by structure models, we find that 
Jupiter's formation location is preferably at an orbital distance of $1\lesssim a \lesssim 10$ au once the accretion of planetesimals dominates.
We find that Jupiter could accrete between $\sim$1 and $\sim$15 $\Mearth$ of heavy elements during runaway gas accretion, depending on the assumed initial surface density of planetesimals and the prescription used to estimate the heavy-element accretion during the final stage of the planetary formation. This would  
yield an envelope metallicity of $\sim$0.5 to $\sim$3 times solar. 
By computing the solid (heavy-element) accretion during the detached phase, we infer a planetary mass-metallicity ($\Mp$-$\Mz$) relation of 
$\Mz \sim \Mp^{2/5}$ when a gap in the planetesimal disk is created, and of $\Mz \sim \Mp^{1/6}$ without a planetesimal gap.}
{Our hybrid pebble-planetesimal model can account for Jupiter's bulk and atmospheric enrichment. The high bulk metallicity inferred for many giant exoplanets is difficult to explain from standard formation models. 
This might suggest a migration history for such highly enriched giant exoplanets and/or giant impacts after the disk's dispersal.}


   \maketitle

\section{Introduction} \label{intro}
Jupiter's formation has been studied extensively since the very early works on planet formation \citep[e.g.,][]{Bodenheimer1986,P96}, and yet, there are still many open questions regarding its origin. 
In the standard model for giant planet formation (core accretion), the formation of a giant planet begins with the accretion of solids followed by accretion of gas disk (see Helled et al. 2014 for review and references therein).
The timescale for core formation depends on the size of the assumed solids. Classical models of core accretion assumed these solids to be large planetesimals with sizes of 100 km \citep{P96, Alibert05}, while more recent studies pointed out the importance of growing a core by cm-size pebbles  \citep{LJ14}. 
In this work we use the terms \textit{solids} and \textit{heavy elements} as synonyms, despite of the physical state of such compounds after being accreted by the protoplanet.
As the protoplanet reaches crossover mass (where the mass of the H-He envelope equals the mass of heavy elements), the gas accretion rate continuously increases and exceeds the heavy-element accretion rate. 
Soon after crossover time,  the disk can no longer supply the gas required by the cooling and strong self-gravity of the protoplanet. 
At this point, the gas accretion rate is limited by the availability of the disk to supply gas, the protoplanet becomes \textit{detached} from the disk, and the gas accretion rate is dictated by the disk hydrodynamics  \citep{Tanigawa02, Lissauer09, Mordasini13}. 

It is still unknown whether giant planet formation is dominated by the accretion of small (pebbles) or large (planetesimals) solids.  
Typically, accretion of planetesimal requires longer core formation timescale due to the moderate accretion rates \citep{Fortier07,Fortier13}. 
On the other hand, pebbles undergo orbital decay through the interaction with the disk of gas, 
and they can be easily and effectively accreted by a growing core, 
due to efficient gas friction within the Hill's sphere \citep[e.g.,][]{Ormel17}. 
This gave to the pebble accretion scenario a special attractiveness to solve the historical problem of planetesimal accretion being too slow to grow a critical core in timescales comparable to the disk lifetime \citep{LJ14, Lambrechts14}.

When a planet grows by accreting pebbles, at certain point it reaches the  the so-called \textit{pebble isolation mass}. This is the necessary mass to perturb the disk and create a pressure bump beyond the orbit of the protoplanet. 
Within a pressure bump the sign of the pressure gradient is reversed, so the direction of the drift of pebbles changes accordingly. Thus, a pressure bump acts as a ``trap" of pebbles, and pebbles cannot be accreted by the protoplanet any longer.
Hence, once the protoplanet attains the pebble isolation mass, pebble accretion is halted.
The pebble isolation mass is estimated to be $\sim$ 20 $M_{\oplus}$ and its exact value depends on the assumed disk properties \citep[e.g.,][]{Lambrechts14, Sareh18, Bitsch18}.

Recent analysis of meteoritic data of isotopic anomalies found in iron meteorites were used by  \citet[][]{Kruijer17} (hereafter K17) to constrain Jupiter's formation chronology. 
According to K17, the isotopic anomalies indicate that carbonaceous and non-carbonaceous chondrites were accreted isolated from each other between times $\sim$1 and  $\sim$3-4 Myr after formation of CAIs (calcium-aluminium-inclusions).  
K17 proposed that proto-Jupiter reached pebble isolation mass at time 1 Myr, and planetesimals were formed at both sides of proto-Jupiter, until time $\sim2$ Myr for the non-carbonaceous chondrites 
and until time$\sim$3-4 Myr for the carbonaceous chondrites. At that time, Jupiter became massive enough ($\sim$ 50 \ME) to scatter planetesimals, reconnecting both reservoirs, and matching in this way, the timescale inferred by the meteoritic record. 

\citet{A18} (hereafter A18), built upon the interpretation of K17, and investigated in detail the validity of this formation scenario, in which between times 1 and 3 Myr from the formation of CAIs, proto-Jupiter separates the solids.
They found that in order to fulfil this time constraint, a solid accretion rate of $\sim 10^{-6}-10^{-5} \Mearth/\rm yr$ is needed to prevent the onset of runaway gas accretion before time $\sim$3 Myr. 
Since the protoplanet reached the pebble isolation mass at time $\sim$1 Myr, after this time pebbles could not be accreted by the planet, and hence, 
the heat source during that period must come from the accretion of planetesimals.

The exact formation process of Jupiter is not fully constrained and it is unknown whether the early stages of the planet's formation are dominated by pebble accretion, planetesimal accretion, or both, as recently suggested by A18. 
In the hybrid formation scenario of A18, there are three formations stages: \\
{\it Phase-1}: a rapid pebble accretion supplied the major part of Jupiter's core mass. In this stage planetesimal accretion is negligible, and large primordial planetesimals are thought to be excited by the growing planet reaching high collision velocities that lead to destructive collisions, and the creation of small, second-generation planetesimals. \\
{\it Phase-2}:  accretion of (small) planetesimals which provides enough energy to hinder runaway gas accretion for $\sim$2 Myr. In this stage Jupiter is massive enough to prevent pebble accretion.\\
{\it Phase-3}: runaway gas accretion, as usual in core accretion models \citep{BP86, P96, Lissauer09}.

In this scenario, phase-2 and phase-3 are somewhat similar to the ones in the traditional planetesimal accretion scenario \citep{P96}, although phase 2 is defined differently since it starts at the \textit{pebble} isolation mass and not at the \textit{planetesimal} isolation mass.
The role of planetesimal/pebble accretion in giant planet formation, and therefore the characteristics of the different formation phases,  is still being investigated.

 Several important questions regarding Jupiter's formation remain unanswered, for example: 

(i) Where did Jupiter's formation start? 

(ii) How many heavy elements can be accreted by the growing planet, and where did they come from? 

(iii) What is the expected structure and global water abundance in Jupiter?

Unfortunately, these questions cannot be answered easily. However, the combination of new data and new theoretical models allows us to better understand Jupiter's origin.
In this work, we investigate Jupiter's growth history accounting for different planetesimal sizes and accretion rates. 
In particular, we focus on heavy-element accretion during the last stage of formation and connect it to Jupiter's structure models as well as to the mass-metallicity relation of giant exoplanets.

\section{Methods}\label{method}
We investigate Jupiter's in situ formation in the framework of the core accretion model. As in A18, we assume that a certain core is already formed by pebble accretion at time =1 Myr of the disk evolution. 
That is the approximate time at which proto-Jupiter reaches the pebble isolation mass and pebble accretion stops (K17, A18).
Our simulations begin at that point, and we assume that from that time on the accreted heavy elements are in the form of large planetesimals. 
During the early stages of Jupiter's growth, the protoplanet is embedded in the protoplanetary disk, and no clear boundary delimits the protoplanet from the disk. This phase is known as the \textit{attached phase}.
However, once the mass of the envelope surpasses the mass of the core, the gas accretion onto the protoplanet becomes so large, that the disk can no longer supply the amount of gas required by the contraction of the protoplanet. At this stage gas accretion is limited by the disk's supply (\textit{detached phase}).

In addition to the attached and detached phases, we will refer throughout the paper to the following three phases of Jupiter's growth, in analogy to A18:\\
\textit{Phase 1} corresponds to the growth of the core dominated by pebble accretion. \\
\textit{Phase 2} starts when the pebble isolation mass is reached, and finishes at the crossover mass. During this phase the solid accretion is dominated by large planetesimals (10-100 km) (Sect.\ref{sec_OG}). 
The protoplanet accretes gas concurrently, at a rate that is dictated by the cooling of the envelope or its Kelvin-Helmholtz timescale (Sect.\ref{sec_GasAtt}).\\
\textit{Phase 3} starts at crossover mass and finishes when the mass of Jupiter is attained. During this phase gas is accreted in a runaway fashion. 
Soon after crossover time, the detached phase \citep{Mordasini13}, or, equivalently, the "the disk-limited accretion" phase \citep{SI08} begins (see Sect.\ref{sec_det}).

\subsection{Planetesimal Accretion}
The initial size of  planetesimals in the Solar Nebula is still a matter of debate \citep{Kenyon12, Weiden11,Morby09}. 
Simulations based on the streaming instability model, the most accepted scenario to explain planetesimal formation, predict initial sizes in the range of $\sim$ 10-100 km \citep[e.g,][]{Simon17} .  
Therefore, in this work we use the more accepted planetesimal size of $r_{\rm p} = 100$ km for the nominal model.  This is in contrast to A18 where small planetesimals were required.  
We also show results for $r_{\rm p} = 10$ km in Sect.~\ref{results_Mz}.

\subsubsection{Oligarchic Growth}\label{sec_OG}
We implement the planetesimal accretion rate given by \citet{Guilera10} and \citet{Fortier13}, and as originally presented by \citet{Inaba01}; which is hereafter referred to as Oligarchic Growth (OG). 
This model includes the gravitational stirring on the planetesimals due to the presence of the protoplanet and the damping of eccentricities by gas drag. An equilibrium state between the stirring and the damping is assumed, 
which is appropriate for the case of large planetesimals as assumed here \citep{Chambers06}.

Under equilibrium, the root mean square eccentricity of planetesimals ($e$) is given by \citep{Thommes03}:
\begin{equation}\label{e_rms}
e = 1.7 \, \bigg(\frac{ m_{\rm p}^{1/3} \rho_{\rm p}^{2/3}}{b \, C_D \, \rho_{\rm gas} \, a_{\rm p}}\bigg)^{1/5} \bigg(\frac{\Mp}{M_*} \bigg)^{1/3} ,
\end{equation}
where $m_{\rm p}$ is the planetesimal's mass,  $\rho_{\rm p}$ its density,  $b$ a constant equal to 10, $C_D $ the non-dimensional drag coefficient taken as 0.7, and $\rho_{\rm gas}$ the gas density.
In the equilibrium approximation, $i = e /2$.

The planetesimal accretion rate is given by \citep{Chambers06}:
\begin{equation}\label{MdotF13}
\Mzdot= \Omega \, \Sigma \,\Rh^2 \, P_{\rm coll} \, ,
\end{equation}
where $\Omega$ is the Keplerian frequency, $\Sigma$ is the surface density of planetesimals, $\Rh$ the Hill radius, and $P_{\rm coll}$ the collision probability between the embryo and the planetesimals.

The surface density of planetesimals evolves with time as a result of accretion onto the protoplanet and scattering. 
The protoplanet accretes planetesimals from a feeding zone of half-width = 5 $\Rh$, and ejects planetesimals at a rate given by \citep{IdaLin04}:

\begin{equation}
\frac{\rm accretion \, rate }{\rm ejection \, rate} = \bigg(\frac{v_{\rm {esc,disk}}}{v_{\rm surf, planet}} \bigg)^4 
\end{equation}
where $v_{\rm esc, disk} = \sqrt{2G \Msun / a}$ is the escape velocity from the central star at the planet's position, and 
$v_{\rm surf, planet} = \sqrt{G \Mp / R_c} $ is the planet's characteristic surface speed; $R_c$ being the  capture radius. 

The collision probability onto the protoplanet, $P_{\rm coll}$ depends on the excitation between the embryo and planetesimals, and is given by:
 \begin{equation}
 P_{\rm coll} = \text{min}(P_{\rm med}, (P_{\rm high}^{-2} + P_{\rm low}^{-2})^{-1/2}) \, ,
 \end{equation}
 where 
 \begin{align}
 P_{\rm high} =&  \frac{(R_{\rm cap}+ r_{\rm p} )^2}{2 \pi \Rh^2}  \bigg( I_F(\beta)   + \frac{6\Rh I_G(\beta)}{(R_{\rm cap}+ r_{\rm p}) {\tilde{e}}^2} \bigg) \, , \label{Phigh} \\
 P_{\rm med} =& \frac{(R_{\rm cap}+ r_{\rm p} )^2}{4 \pi \Rh^2 \tilde{i}}  \bigg( 17.3 + \frac{232 \, \Rh}{R_{\rm cap} + r_{\rm p}}  \bigg) \, , \label{Pmed}\\
 P_{\rm low}  =& 11.3 \, \bigg( \frac {R_{\rm cap}+ r_{\rm p}}{\Rh} \bigg)^{1/2}, \label{Plow}
 \end{align}
 where $\tilde{e} = a e / \Rh$, $\tilde{i} = a i / \Rh$, and $\beta = \tilde{i} / \tilde{e} $. 
 $R_{\rm cap}$ is the capture radius of the protoplanet, i.e, the effective cross section of collision when a core is surrounded by a gaseous atmosphere. The capture radius reduces to the core radius in the limit of no atmosphere.
 For the computation of the capture radius, we implement the prescription given by \citet{InabaIkoma03}.
 
 The functions $I_F(\beta)$ and $I_G(\beta)$ are approximately given by \citep{Chambers06}:
\begin{align}
I_F(\beta) =& \frac{1 + 0.95925 \, \beta + 0.77251 \, \beta^2}{\beta \, (0.13142 + 0.12295 \, \beta)}, \\
I_G(\beta) =& \frac{1 + 0.3996 \, \beta}{ \beta \, (0.0369 + 0.048333 \, \beta + 0.006874 \, \beta^2) }.
\end{align}

\subsubsection{Detached Phase}\label{SI08_methods} 
For our nominal model, we use Eq.\ref{MdotF13} during all the growth of Jupiter.  
However, that accretion rate of planetesimals was calculated for Earth-mass embryos embedded in a gaseous disk. 
When the planet's growth is dominated by the accretion of gas, \citet[][hereafter SI08]{SI08} showed, from analytical fits to N-body simulations, that the accretion rate of planetesimals is given by two equations, depending on the competition between the expansion of the feeding zone and the damping of eccentricity of planetesimals within it.
\par

When the expansion of the feeding zone dominates, planetesimals enter continuously to the protoplanet's feeding zone, so no planetesimal gap is created. 
\citet{SI08} express this condition in terms of $\eta = v_{\rm H}/v_{\rm damp}$, where   $v_{\rm H}$ is the rate of expansion of the feeding zone and $v_{\rm damp}$ the rate of damping of eccentricity by gas drag.
The non-gap case fulfils $\eta > 1$, and in this case the accretion rate of solids is given by \citep[Eqs.17 and 24 of][]{SI08}:
\begin{equation}\label{nogap}
\MzdotNG=  10^{-6} \, {\rho}^{1/2} \, \bigg(\frac{\Rp}{\RE}\bigg)^2  f_d  \, \bigg(\frac{v_{\rm H}}{v_{\rm scat}}\bigg)^{0.8}  \Mearth / {\rm yr} , 
\end{equation}
where $\rho$ is the mean density of the protoplanet, $\Rp$ is the physical radius of the protoplanet, $f_d$ is the the surface density of solids in Minimum Mass Solar Nebula (MMSN) units, and $v_{\rm scat}$ is the rate at which planetesimals are scattered from the feeding zone.
The ratio of the speed of Hill's radius expansion and planetesimal scattering follows \citep[][their Eq.22 ]{SI08}: 
\begin{equation}\label{vH_vscat}
\frac{v_{\rm H}}{v_{\rm scat}} = 4.1 \, \bigg({\frac{a}{5 \, \rm au}} \bigg)^{3/2} \bigg({\frac{\Mp}{\Mearth}}\bigg)^{-1/3}  \bigg({\frac{\tau_g}{10^4 \rm yr}}\bigg)^{-1}  ,
\end{equation}
where $\tau_g$ is the timescale to accrete gas, given by $\tau_g = \Mp /\MdotHHe $.

When $\eta < 1$, a planetesimal gap is formed and the accretion rate of solids follows  \citep[Eqs.17 and 25 of][]{SI08}:
\begin{equation}\label{gap}
\MzdotG=  10^{-6} \,  {\rho}^{1/2} \, \bigg(\frac{\Rp}{\RE}\bigg)^2  f_d  \, \bigg(\frac{v_{\rm H}}{v_{\rm damp}}\bigg)^{1.4}  \Mearth / {\rm yr} ,   
\end{equation}
where:
\begin{equation}\label{vH_vdamp}
\frac{v_{\rm H}}{v_{\rm damp}} = 0.8 \, \bigg({\frac{a}{5 \, \rm au}} \bigg)^{3/4} \bigg({\frac{\Mp}{\Mearth}}\bigg)^{-1/6}  \bigg({\frac{\tau_g}{10^4 \rm yr}}\bigg)^{-1}    \bigg({\frac{\tau_{\rm damp}}{10^4 \rm yr}}\bigg)^{1/2}  ,
\end{equation}
$\tau_{\rm damp}$ is the timescale to damp the planetesimals' eccentricity, and corresponds approximately to $\tau_{\rm damp} \sim 10^6$ yr for large planetesimals as the ones considered in this work \citep{SI08}.
Along this work, we refer to the implementation of Eqs.\ref{nogap} to \ref{vH_vdamp} as the SI08 prescription.
Note that equations \ref{nogap} and \ref{gap} are the general form of the fit found by SI08, and not the particular case of $\rho= $1 \rhocgs \, assumed in Eqs. 24 and 25 of SI08. 
We test the effect of these accretion rates during the runaway gas phase, when the planet is detached from the disk. The results are presented in Sect.\ref{MMrel}.

Note that the above solid accretion rates depend on the planetary radius as $\sim \Rp^{1/2}$.
1D simulations have shown that during the detached phase the planetary radius contracts very fast to $\sim$ 2 Jupiter radius, and remains fairly constant during the remaining of gas accretion  \citep{Lissauer09, Mordasini13}.
Following those studies, we adopt $\Rp = 1.8 \, R_{\rm Jup}$ when implementing Eqs.\ref{nogap} and \ref{gap}.

\subsection{Disk Model}\label{diskB15}
Initially, giant planets grow while being embedded in a protoplanetary disk. The disk temperature and pressure affect the outer boundary of the protoplanet, and hence, a disk model is required to study the planetary growth in a consistent manner.
We adopt the disk model from \citet[][hereafter B15]{Bitsch15a}. 
This disk evolution model adopts the accretion rate $\dot{M}$ onto the central star derived from observations \citep{Hartmann98}:

\begin{equation}\label{Hartmann}
\log \left( \frac{\dot{M}}{M_\odot/ {\rm yr}} \right) = -8.00 -1.40 \times \log \left( \frac{t + 10^5 {\rm yr}}{10^6 {\rm yr}} \right) ,
\end{equation}
where $t$ is the evolution time, and the error bars from the original expression in \citet{Hartmann98} are neglected.

$\dot{M}$ is related to the disk's viscosity  and the gas surface density $\SigmaG$ via:
\begin{equation}\label{sigma_gas}
\dot{M} = 3 \pi \nu \SigmaG = 3 \pi \alpha H^2 \Omega \SigmaG ,
\end{equation}
 where $H$ is the disk's scale height, and $\Omega$ the Keplerian frequency.
$\alpha$ represents the viscosity and only acts as a heating parameter. B15 adopted $\alpha = 0.0054$ to perform the radiative-hydrodynamical simulations, so this is the value that must be assumed for consistency. 
After computing the temperature profile from the formulas presented in the Appendix of B15, the disk scale height can be obtained from:
\begin{equation}
T = \bigg( \frac{H}{r} \bigg)^2 \, \frac{GM_*}{r} \frac{\mu}{R}  .
\end{equation}
Finally, $\SigmaG$ can be calculated from Eq.\ref{sigma_gas}. 

The initial surface density of planetesimals (at time =1 Myr), $\Sigma_1$, is a free parameter of the model. 
Note that this quantity is typically a free parameter in planet formation simulations, and is hidden in the assumed initial disk dust-to-gas ratio which is often set to $\sim$0.01 guided by observations. 
However, studies of dust growth and drift \citep[e.g,][]{Birnstiel12} suggest that the initial dust-to-gas ratio can vary significantly from the planetesimal-to-gas ratio in a given location of the disk \citep{Drazkowska16}. 
In addition, planetesimals can be formed in regions where dust is piled up, and the dust-to-gas ratio reaches values of $\sim$1. 
In fact,  the planetesimal-to-gas ratio could be even larger than 1 at t=1 Myr of the disk's evolution, a few 10$^5$ years after the onset of planetesimal formation  \citep[][from their Figs.2 and 4]{Drazkowska16}. 
Since various theoretical models and observations reveal  an unprecedented complexity in the dust distribution within protoplanetary disks, 
the simplification of assuming an initial planetesimal-to-gas ratio equal to a fixed dust-to-gas ratio is no longer justified and is clearly an over-simplification.

\subsection{Gas accretion: Attached Phase}\label{sec_GasAtt}
We simulate the planetary growth  by solving the standard planetary internal structure equations, 
assuming the luminosity  results from the accretion of solids and envelope contraction  \citep[see][for details]{Alibert13, Venturini16}:  
\begin{eqnarray}
\frac{d r^3}{d m} & = & \frac{3}{ 4 \pi \rho}, \\
\frac{d P}{d m}&  = & \frac{- G ( m + M_{\rm core} ) }{4 \pi r^4} , \\
\frac{d T}{d P}&  = & {\rm min}(\nabla_{\rm conv}, \nabla_{\rm rad}),
\end{eqnarray}
where $G$ is the gravitational constant, and $r,P,T, \rho$ are the radius, the pressure,
temperature and density inside the {envelope}, respectively. 

The density is given as a function of $T$ and $P$ by the equation of state (EOS). 
For the nominal case, we assume that the solids reach the core, meaning that the envelope keeps a  H-He composition. 
For this case we use the EOS of \citet{SCVH}. Since we assume that large planetesimals are accreted, it is expected that despite of thermal ablation, the material is deposited deep in the envelope, especially the refractory materials at early formation stages \citep[e.g.,][]{Valletta19}. 
We also consider an enriched-envelope case in which we assume that all the volatiles remain in the envelope and mix uniformly throughout it \citep{Venturini16}. 
Although this case is probably more realistic for pebble accretion, we still evaluate this possibility in \ref{envelope_composition}.
We suggest that future formation models should couple the heavy element deposition with the resulting compositional gradient and its effect on the internal structure. 

The temperature gradient is given by either the radiative gradient ($\nabla_{\rm rad}$): 
\begin{equation}
\nabla_{\rm rad} = \frac{ 3 \kappa L }{ 64 \pi \sigma G ({m + M_{\rm core}}) T^3}
\end{equation}
or the convective gradient, equal to the adiabatic one. In these formulas, 
 $\sigma$ is the Stefan-Boltzmann constant, $\kappa$ is  the opacity, and $\Mcore$ is the mass of the core.
 
 For the opacity, we take $\kappa = \text{max} (f \,\kappa_{\rm dust}, \kappa_{\rm gas})$, where the dust opacity is taken from \citet{BL94}, the gas from \citet{Freedman14}, and for the nominal model we take $f=1$.
 For a solar composition this definition is equivalent to consider the \citet{BL94} opacities throughout the envelope, but with the current definition we can scale down the dust opacity to consider grain growth. 
 Moreover, the \citet{Freedman14} gas opacities account for a large range of envelope metallicities, so they are  better suited for the cases where envelope enrichment is considered during growth.
 
The internal structure equations are solved, using as boundary conditions the pressure and temperature in the protoplanetary 
disk at the position of the planetary embryo and defining the planetary radius as a combination of the Hill and Bondi 
radii \citep{Lissauer09}: 
\begin{equation}
	R_{\rm P} = \frac{G \Mp}{ \left( C_{\rm S}^2 + 4 G \Mp / \Rh \right) } 
\end{equation}
where $C_{\rm S}^2 $ is the square of the sound velocity in the protoplanetary disk at the planet's location $a$,
and the Hill radius $\Rh = a \left( \frac{ \Mp}{3 M_\odot } \right)^{1/3}$. 


\subsection{Gas accretion: Detached Phase} \label{sec_det}
Once the gas accretion given by the gas disk dynamics is smaller than the gas required from the protoplanet by its own cooling, the protoplanet continues accreting gas at a rate determined by the disk flow ($\dot{M}_{\rm gas, disk}$). In other words, the gas accretion onto the protoplanet is given by:
\begin{equation}
	\dot{M}_{\rm gas} = \text{min} \, \{\dot{M}_{\rm gas, KH}, \dot{M}_{\rm gas, disk}\}
\end{equation} 
where $\dot{M}_{\rm gas, KH}$ is the gas accretion obtained from computing the cooling of the protoplanet (KH standing for Kelvin-Helmholtz), as explained in Sect.\ref{sec_GasAtt}, and  $\dot{M}_{\rm gas, disk}$ is the accretion rate of gas that the disk is able to supply to the protoplanet. For the latter, we take Eq.(3) from \citet[][hereafter L09]{Lissauer09}:

\begin{equation}\label{L09}
	\log \Bigg( \frac{\Mdothydro}{\SigmaG a^2  /P}  \Bigg)  =  c_0 + c_1 \log \Bigg( \frac{\Mp}{M_*} \Bigg) + c_2 \log^2 \Bigg( \frac{\Mp}{M_*} \Bigg), 
\end{equation}
where $P$ being the planet's orbital period, $c_0 = -18.67$, $c_1 = -8.97$ and $c_2=-1.23 $. This expression was obtained as a fit to 3D hydrodynamical simulations for $\alpha = 4 \times 10^{-3}$. 
Following also L09, we multiply the gas accretion rate by a linear function equal to 1 at 0.85 Jupiter mass and equal to 0 at 1 Jupiter mass ($1 M_J$), in order to terminate gas accretion when reaching one Jupiter mass. 
It should be noted that in our simulations we do not consider the accretion shock during the last phase of rapid gas accretion which can lead to entropy gradients inside the planet and persist for  up to 10 Myr after its formation \citep{Cumming18}. 
While this process is highly important for the determination of Jupiter's primordial entropy and heat transport mechanism at early ages, it is less likely to affect the heavy-element accretion during the last stage of formation. This, however, should be investigated in detail in future studies. 

A summary of the main recipes and parameters used in this work is shown in Table \ref{tab_glossary}.

\begin{table*}
\caption{Glossary of main parameters and recipes used in this work.} 
\label{tab_glossary}      
\centering                                      
\begin{tabular}{l l l l }          
\hline                      
\hline
Symbol of parameter/recipe  & Meaning & Section where & Used in simulations / Results shown in /\\    
& & it is introduced  & Other comments \\
\hline                                   
$\Sigma_1$ & Initial surface density of planetesimals & \ref{diskB15}  & value at t = 1 Myr of disk evolution   \\ %
$\Mcore$$_{,1}$ &  Initial core mass  & \ref{sec_results} &  value at t = 1 Myr of disk evolution \\ %
\hline
OG & Oligarchic Growth & \ref{sec_OG} & all \\ 
SI08 & \citet{SI08} & \ref{SI08_methods}  & all, once the detached phase starts \\
 $\dot{M}_{\rm gas, disk}$ & Gas accretion given by disk's supply  & \ref{sec_det} & all, once the detached phase starts \\
\hline
$a$ & Planet location & \ref{sec_a1} & nominal value: a = 5 au, \\
&&&other adopted values: a = 1,3,10 au  \\
&&& Figs.\ref{maps_a1}, \ref{MMSN}, Tables \ref{table_Mz}, \ref{table_Mz_MMSN} \\ 
 \hline
$\rp$ & Planetesimal size& \ref{results_Mz} & 100 km for all except \\
& & & in Fig.\ref{maps10km} where is set to 10 km \\
\hline

Reduced opacity & Dust opacity reduced by a factor 10 & \ref{envelope_composition} & Fig.\ref{kappa01} and Table \ref{table_Mz} \\
Envelope enrichment & Variable envelope metallicity due to ices' sublimation & \ref{envelope_composition}  & Fig.\ref{kappa01} and Table \ref{table_Mz} \\

\hline                                            
\end{tabular}
\end{table*}


\section{Results} \label{sec_results}
\subsection{Formation at $a=5$ au}

\begin{figure}
\begin{center}
	\includegraphics[width=\columnwidth]{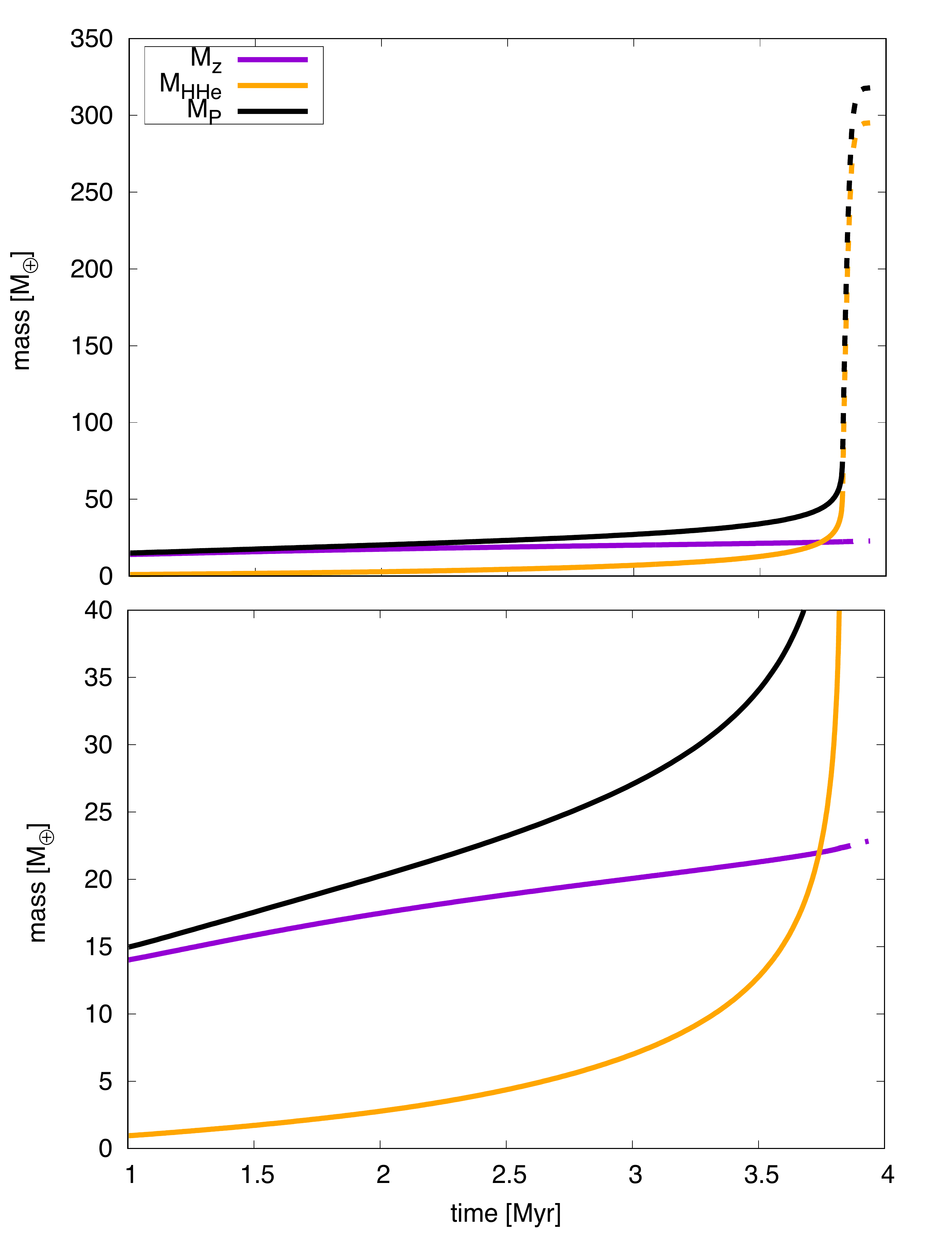}
	\caption{An example of planetary growth, where the initial surface density of planetesimals is $\Sigma_1$ = 7 \gcmcuad \, and the initial core mass is $\Mcore$$_{,1}$ =  14 \ME. $M_Z$ is the mass of heavy elements, $\MHHe$ the mass of H and He, and $\Mp$ the total mass of the planet. The solid lines indicate that the protoplanet is in the attached phase, whereas the dotted lines show the protoplanet in the detached phase. The bottom panel is simply a zoom to appreciate better the growth of $M_Z$.}
\label{fig1}
\end{center}
\end{figure}

To illustrate the output of the model, we first show, in Fig.\ref{fig1}, an example of planetary growth at $a=5$ au that matches the meteoritic constraints. That is, Jupiter attains pebble isolation mass at time 1 Myr and $\Mp$ = 50 $\Mearth$ at a time of 3-4 Myr of disk evolution (K17). 
For this particular case, the initial core mass is $\Mcore$$_{,1}$ = 14 \ME, and the initial planetesimal surface density is $\Sigma_1$ = 7 \gcmcuad \,. The planetesimal size assumed is  $\rp$= 100 km, and the planetesimals' density, $\rhop $= 1 \rhocgs.
The solid lines indicate the growth within the attached phase, and the dotted lines within the detached phase. The crossover mass is $\Mcross$ = 21.9 \ME\, with the crossover time being $\tcross$=3.7 Myr. 
The detached phase starts when the total planetary mass is $\Mp = 82 \, \Mearth$, the mass of heavy elements is $\Mz$ = 22.3 \ME, and the accretion rate of H-He is $\MdotHHe = 9.7 \times 10^{-3} \Mearth $/yr. 
During the detached phase, additional 0.6 \ME \ of heavy elements are accreted (OG prescription).

\subsubsection{Meteoritic constraints and comparison with A18}

\begin{figure*}
\begin{center}
	\includegraphics[width=\textwidth]{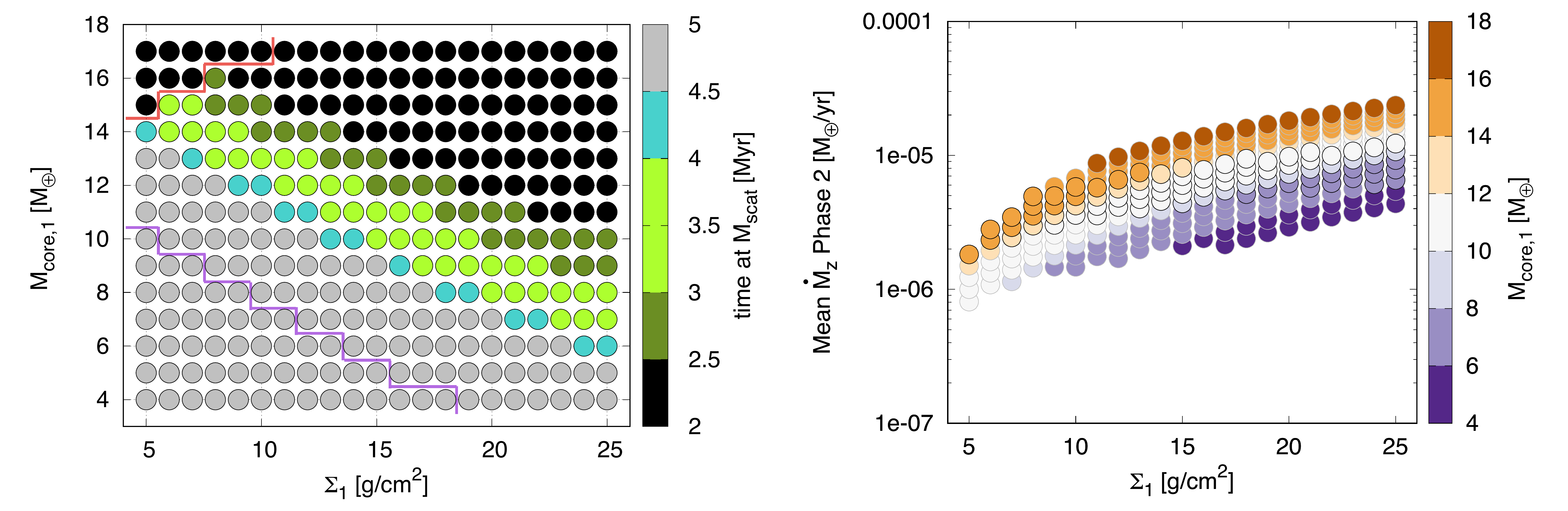}
	\caption{Grid of simulations for  $a$= 5 au and $\rp$ = 100 km. \textit{Left:} Mean time to reach the scattering mass as a function of the initial surface density of planetesimals ($\Sigma_1$) and the initial core mass ($\Mcore$$_{,1}$). The light green color indicates the cases compatible with the meteoritic record. For grey circles the growth is too slow (longer than 4.5 Myr), and for black ones, too fast (shorter than 2.5 Myr). Points in the upper left corner above the red line are supercritical and their structure cannot be computed. Points in the lower left corner below the purple line correspond to cases whose growth is so slow that the mass of Jupiter is not reached within 10 Myr of disk evolution. \textit{Right:} Mean accretion rate of solids during Phase 2 as a function of  the initial surface density of planetesimals. The initial core mass is shown in the color-bar. Circles with black border represent cases where the meteoritic constraints are fulfilled (green band of left panel).}
\label{maps_nominal}
\end{center}
\end{figure*}

Figure \ref{maps_nominal} summarizes the results for all simulations at $a$ = 5 au with planetesimal sizes of 100 km. We refer to these set of simulations as the \textit{nominal case}. Each point in each panel is the growth of one planet, characterized by an initial core mass ($\Mcore$$_{,1}$, the core mass at $t$ =1 Myr), and planetesimal surface density at 1 Myr of disk evolution ($\Sigma_1$). The left panel plots the time required to reach $\Mscat$, the mean mass the protoplanet must attain to scatter planetesimals into orbits inside proto-Jupiter, which range between 30 and 70 $\Mearth$ \citep{A18}.
The green circles represent the models that fulfil the meteoritic constraint of reaching $\tscat \approx3-4$ Myr after CAI formation. 
The black circles indicate a too fast formation timescale, of $\tscat < 2.5$ Myr; and the opposite for grey ones, with $\tscat$ > 4.5 Myr. 
Models in black in the upper left corner, above the red line, correspond to those whose $\Mcore$$_{,1}$ is supercritical, and therefore, the growth cannot be computed. These cases would yield too short formation timescales. 

Regarding the trends of the plot, for a fixed $\Sigma$$_1$, the larger the initial core mass, the less time is needed to reach the crossover mass, so the overall formation timescale shortens when we increase $\Mcore$$_{,1}$.  When we move in the horizontal direction, the larger the initial planetesimal surface density, the larger the accretion rate of solids, so the formation timescale also shortens in this direction. We find that when accreting planetesimals of 100 km size at $a= 5$ au, the pebble isolation mass must range between 6 and 16 \ME \, to fulfil the K17 constraints.

Note that at first sight, the values considered for $\Sigma_1$ might look too large. However, in terms of planetesimal-to-gas ratio, $\Sigma_1=25$ \gcmcuad \, corresponds to $\Sigma_1/ \SigmaG=0.231$ in our model, value expected in regions where planetesimals are formed \citep{Drazkowska16}. The ranges of $\Sigma_1$ with its corresponding planetesimal-to-gas ratio  and equivalent in units of MMSN are listed in Table \ref{tab_2}.
The right panel of the figure shows the mean accretion rate of solids during Phase-2 (between time 1 Myr and crossover time). We corroborate that the values range between $\Mzdot \approx 10^{-6}-10^{-5} \Mearth$/yr, as found by A18. 

The timescale for Jupiter's formation provided by the meteoritic constraints can be matched when the second stage of Jupiter's growth is dominated by the accretion of 100 km planetesimals.  This was not found by the simple estimates of accretion rates given in A18. The difference is that in this work, we follow the planet's growth consistently, accounting for the enhancement factor of the accretion due to the presence of an atmosphere. Indeed, the capture radius can be larger than the core radius by a factor of $\sim 20$. This is shown in detail in Appendix \ref{app_Rcap}.  
A large capture radius increases the planetesimal accretion rate (Eqs.\ref{Phigh}-\ref{Plow}), relaxing the implications on the minimum size of planetesimals that match the K17 scenario. 
In summary, we find that the "meteoritic time constraints" can be easily matched by accreting large planetesimals, which are presumably the most abundant type when planets are forming \citep[e.g.,][]{JohansenPPVI}.

\begin{table}
\caption{Values of the initial surface density of planetesimals (in \gcmcuad \, and in units of the MMSN), and  initial planetesimal-to-gas ratio at $a= 5$ au.} 
\label{tab_2}      
\centering                                      
\begin{tabular}{c c c}          
\hline                      
$\Sigma_1$ [\gcmcuad]  & $\Sigma_1 / \Sigma_{\rm MMSN}$ & $\Sigma_1/\SigmaG$ \\    
\hline                                   
1  & 0.365 & 9.237 $\times 10^{-3}$   \\ %
5 & 1.827 & 4.619 $\times 10^{-2}$ \\ %
10 & 3.654 & 9.237 $\times 10^{-2}$ \\ %
15 &  5.481 & 0.139  \\ 
20 & 7.307 & 0.185 \\
25 &  9.134 &  0.231 \\
\hline                                             
\end{tabular}
\end{table}

\subsubsection{Heavy elements in Jupiter: comparison with Jupiter's structure models}\label{results_Mz}

\begin{figure*}
\begin{center}
	\includegraphics[width=\textwidth]{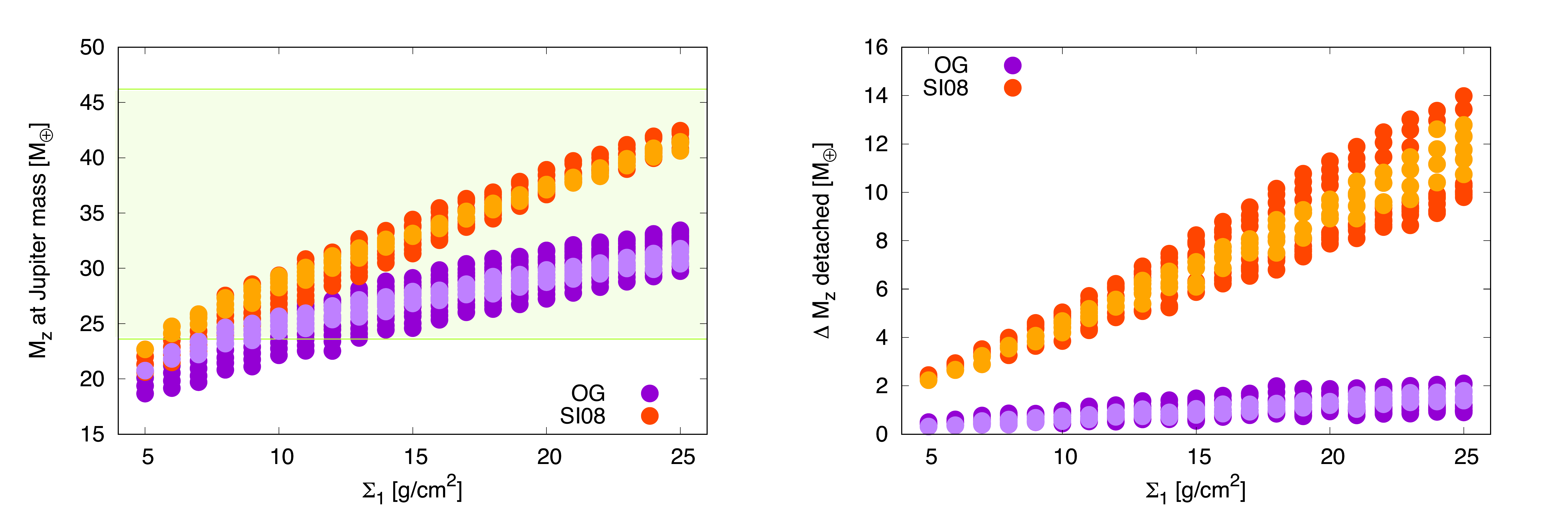}
	\caption{ Simulations for  $a$= 5 au and $\rp$ = 100 km with the same range in ($\Mcore$$_{,1}$, $\Sigma$$_1$) as in Fig.\ref{maps_nominal}. Orange circles correspond to the solid accretion given by SI08, purple circles to Oligarchic Growth (OG). The darker colors represent all runs  that reach Jupiter mass within 10 Myr of disk evolution. The lighter colors highlight the cases that match the meteoritic constraints of K17.  
	\textit{Left:} Total amount of heavy elements accreted until reaching Jupiter mass as a function of the initial surface density of planetesimals ($\Sigma_1$). The green shaded area indicates the values of $M_Z$ compatible with estimates from \textit{Juno} \citep{Wahl17}. 
	\textit{Right:} Amount of heavy elements accreted during the detached phase with the different solid accretion prescriptions. }
\label{heavies}
\end{center}
\end{figure*}

The left panel of Fig.~\ref{heavies} shows the total amount of heavy elements when the protoplanet reaches 1 Jupiter mass (M$_{\rm Jup}$). 
We find that Jupiter could have accreted between 18 and 34 \ME\, of heavy elements during its formation when implementing the OG prescription; and between 20 and 43 \ME\, when implementing the prescription of SI08.
These findings are in good agreement with estimates from recent structure models of Jupiter that fit the new gravity data measured by the \textit{Juno} spacecraft \citep{Wahl17} (green shaded area in the left panel).
Note that the total heavy-element mass accreted until reaching Jupiter's mass is strongly dependent on $\Sigma_1$.

The amount of heavy elements accreted during the final phase of Jupiter's formation (the detached phase) is shown in right panel of figure \ref{heavies}.
The range of heavy elements accreted in this fast stage varies between 0.3 and 14 \ME. 
It should be noted that our results on Jupiter's enrichment during the different growth phases, are independent of the timescale constraint for Jupiter's formation as presented by K17. 
At the same time, our estimates for the total heavy-element mass in the planet rely on prescriptions presented by other studies. Determining the heavy-element mass accreted during Jupiter's formation is challenging and is still a topic of intensive investigation. 
A more robust estimate would be possible only when the different approaches and methods (i.e., quasi-hydrostatic growth, hydrodynamics, N-body interactions) are combined in one theoretical framework. 
Finally, the heavy-element mass estimates from Jupiter structure models also suffer from uncertainties. Our study demonstrates how different assumed parameters affect the inferred composition. This can then be used to better understand the origin of Jupiter's enrichment and for the interpretation of the formation paths of enriched giant exoplanets, as we discuss below.

\begin{figure*}
\begin{center}
	\includegraphics[width=\textwidth]{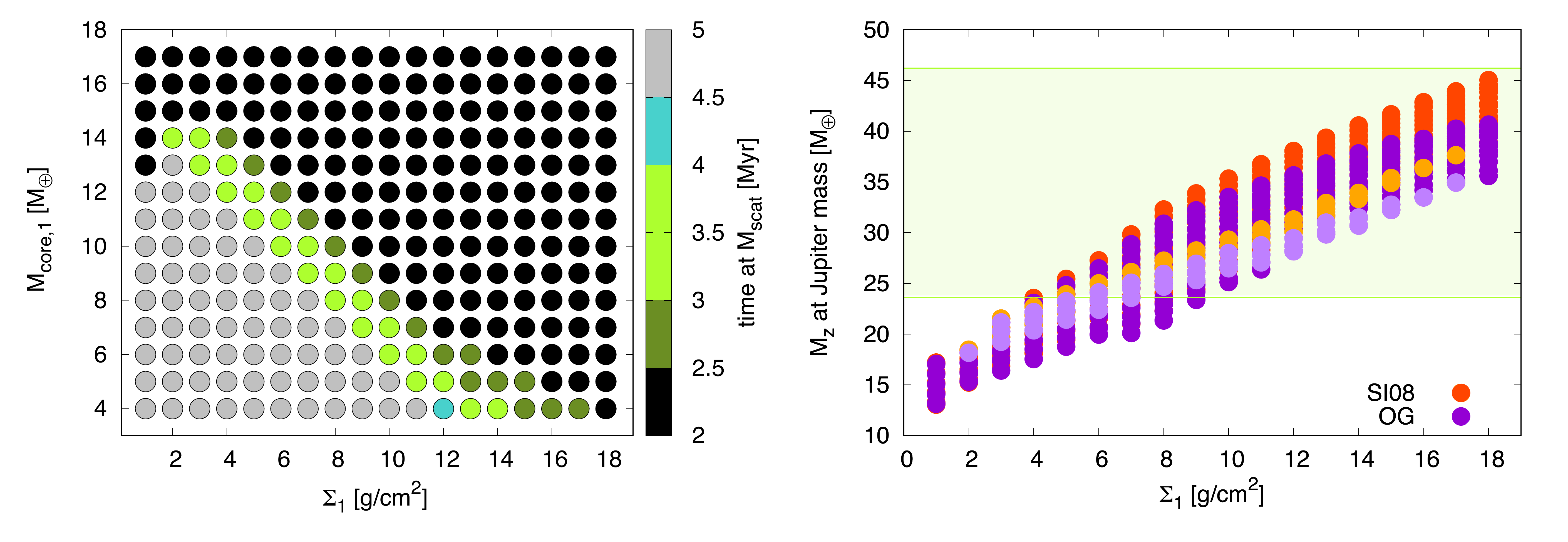}
	\caption{ Grid of simulations for  $a$= 5 au and $\rp$ = 10 km. \textit{Left:} Mean time to reach the scattering mass as a function of the initial surface density of planetesimals ($\Sigma_1$) and initial core mass ($\Mcore$$_{,1}$). The light green color indicates the cases compatible with the meteoritic record. For grey circles the growth is too slow, and for black ones, too fast. \textit{Right:} Total amount of heavy elements accreted until reaching Jupiter mass as a function of the initial surface density of planetesimals ($\Sigma_1$). Transparent circles: all results. Solid circles: cases compatible with the meteoritic constraints.The green shaded area indicates the values compatible with estimates from \textit{Juno} \citep{Wahl17}.  }
\label{maps10km}
\end{center}
\end{figure*}

We next repeat the simulations assuming a planetesimal size of $\rp = 10$ km. The results are shown in figure \ref{maps10km}. 
Compared to the runs of $\rp = 100 $ km, the growth is more efficient, meaning with smaller $\Sigma_1$, the same growth timescale than with $\rp=100$ km can be attained. This is a typical feature of oligarchic growth: smaller planetesimals are more easily accreted \citep[e.g,][]{Fortier07, Guilera10}.

Interestingly, the width of the solutions that fit the scenario of K17 (green band of top left panel of figure \ref{maps10km}) is narrower in this case.
Regarding the solids accreted with $\rp = 10$ km until the planet reaches Jupiter mass, we note that the values are very similar to the case of $\rp = 100$ km (Fig.\ref{maps10km}, right panel), 
but note that the range of $\Sigma_1$ was assumed smaller for this set of simulations.

The full summary of all heavy elements accreted in each phase for all the models is shown in Appendix \ref{app_summary_all_phases}.

\subsubsection{Dependence on Opacity and Envelope Composition}\label{envelope_composition}

\begin{figure}
\begin{center}
	\includegraphics[width=\columnwidth]{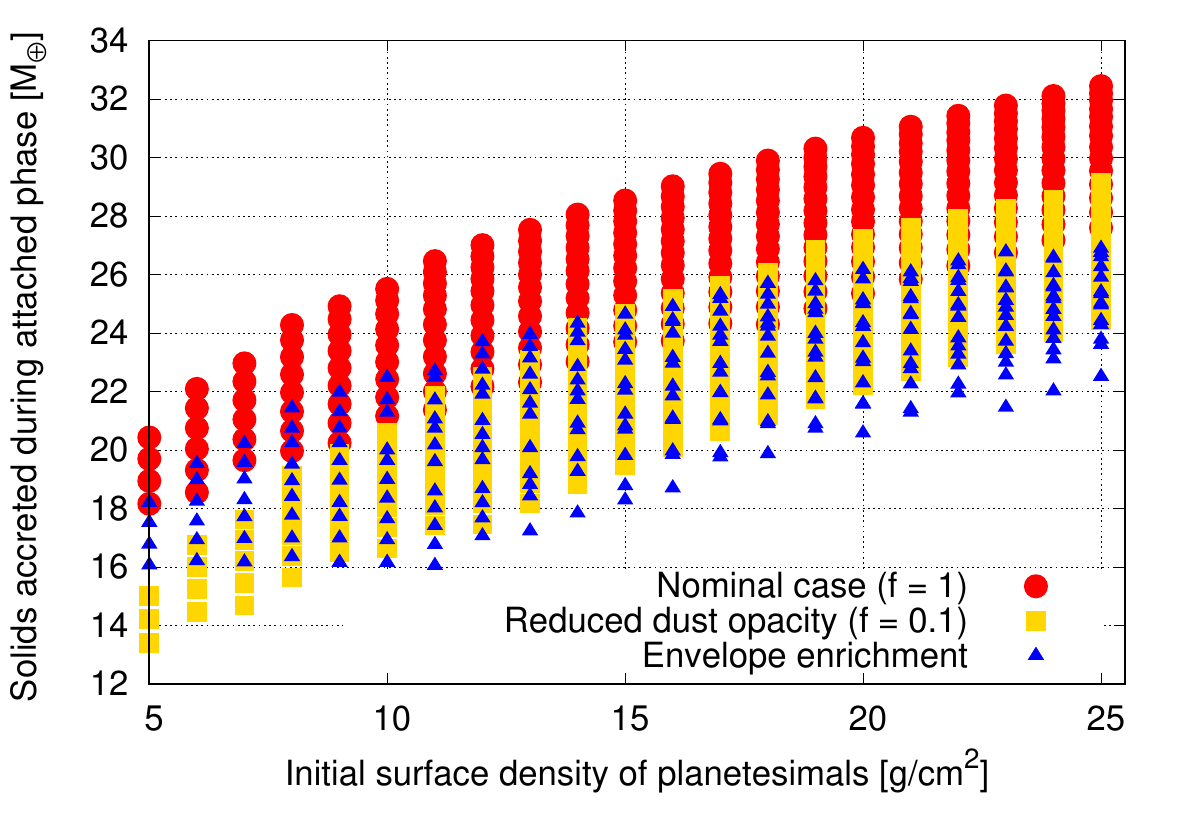}
	\caption{Amount of heavy elements accreted during the attached phase (including the initial core formed by pebble accretion) as a function of the initial surface density of planetesimals for different values of dust opacities (circles and squares) and considering or not envelope enrichment (triangles and circles). 
	All the cases that reach Jupiter mass within 10 Myr of disk evolution are shown. All cases correspond to planetesimal sizes of  $\rp$ = 100 km.
	Red circles indicate the nominal simulations ($f=1$) , squared yellow points represent the case with dust opacity reduced by a factor of 10 ($f = 0.1$), and blue triangles indicate the case where envelope enrichment is included.}
\label{kappa01}
\end{center}
\end{figure}

We next study the dependence of the accreted heavy-element mass on the assumed envelope's opacity. 
The dust opacity could be reduced compared to ISM values due to grain growth and settling \citep[e.g.,][]{Movshovitz10}.
Therefore, we first analyse results for the case where we reduce the dust opacity by a factor of 10 (i.e., $f = 0.1$). 

Figure \ref{kappa01} shows the total amount of heavy elements accreted during the attached phase as a function of the initial surface density of planetesimals, for the simulations that fulfil the meteoritic constraints.
The squares show the results for $f=0.1$ while the circles represent the nominal run ($f=1$). 
We note that for all runs, the amount of solids accreted in the low opacity case is smaller compared to the nominal case, by a shift of $\sim 4 $ \ME.
The fact that the protoplanet accretes less solids when attaining the same total mass is expected: the lower the opacity, the faster the gas accretion \citep{Ikoma00, Movshovitz08}. 
In other words, if the opacity is lower, a smaller core can hold a more massive atmosphere in hydrostatic equilibrium such that the total mass of the planet is the same as in the high opacity case. 
Therefore, at Jupiter mass, for the low opacity case, less solids are expected to be accreted.

A very similar behaviour is observed when envelope enrichment is included.
We run simulations for the case where we assume that half of the solids (the ices) remain homogeneously mixed in the envelope, while the refractories are assumed to sink to the core 
\citep[same set up as in][]{Venturini16, Venturini17}. 
Since in this case the timescale of formation is shorter, we find that when envelope enrichment is considered, the amount of solids accreted by Jupiter is reduced compared to the nominal case (triangles in Fig.\ref{kappa01}).

Overall, the total amount of solids accreted in the three cases is in the range of 14-32 \ME, and the difference between the nominal case and the other two cases are small, of the order of 2-6 \ME.  
Therefore, we conclude that the effect of envelope composition does not affect significantly the estimates of the heavy-element mass accreted by Jupiter.

Interestingly, when envelope enrichment is taken into account, the relation between the amount of heavy elements accreted during the attached phase and the initial planetesimal surface density of solids is flatter than in the nominal case (triangles in Fig.\ref{kappa01}). 
This weaker dependence between $\Sigma_1$ and the amount of accreted heavy elements in the case when envelope enrichment is accounted for, is important, because it would indicate a weak relation between star metallicity and planet metallicity. 
Such a weak relation seems to be present in giant exoplanets, as is shown in Fig.8 of \citet{T16}.

\subsection{Dependence on the formation location}\label{sec_a1}

\begin{figure*}
\begin{center}
	\includegraphics[width=0.9\textwidth]{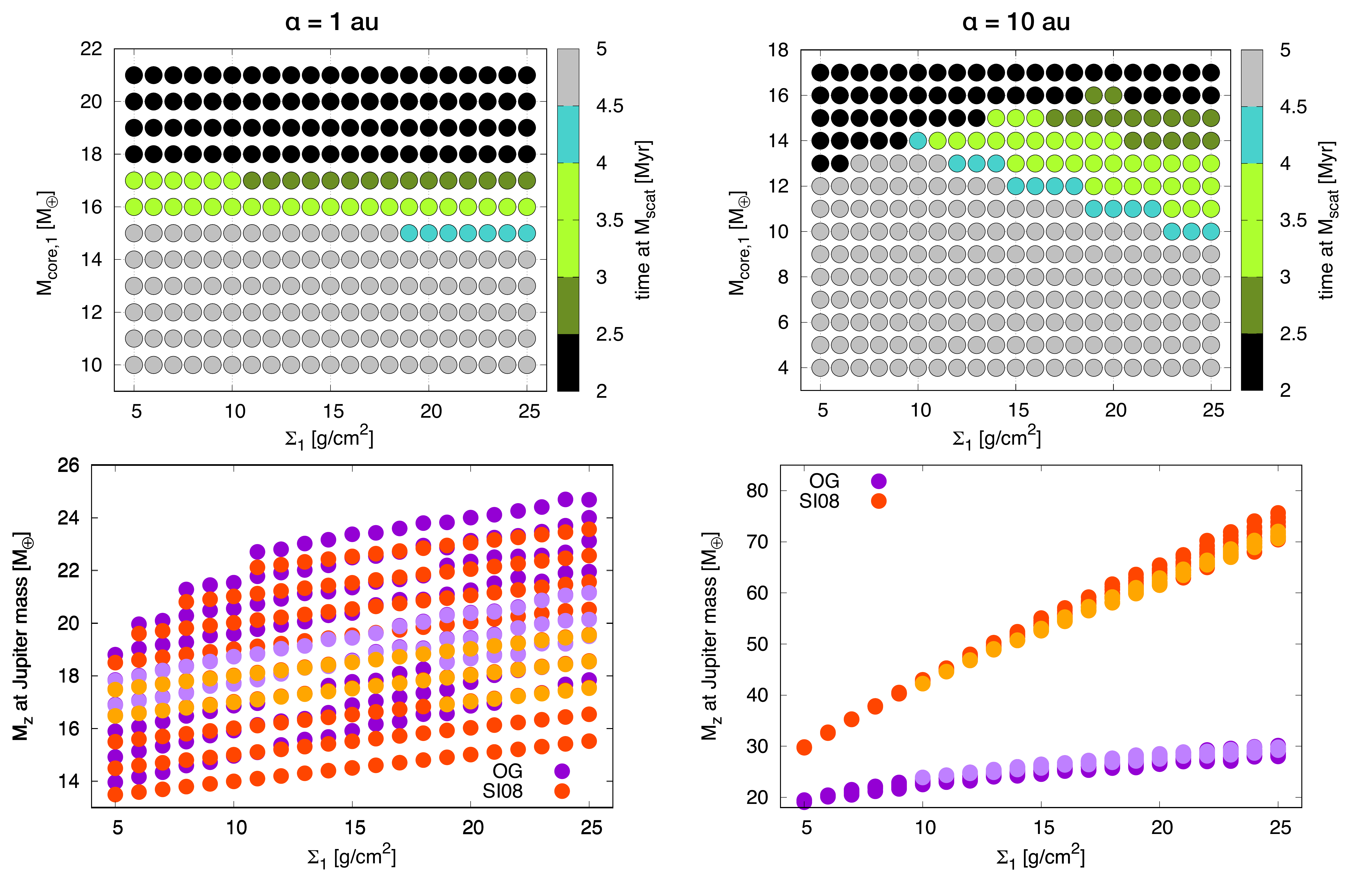} 
	\caption{ Grid of simulations for $a= 1$ au (left) and $a= 10$ au. $\rp$ = 100 km. \textit{Top:} Mean time to reach the scattering mass as a function of initial surface density of planetesimals and initial core mass. The light green color indicates the cases compatible with the meteoritic record. For grey circles the growth is too slow, and for black ones, too fast.
	\textit{Bottom:} Total amount of heavy elements acquired by the planet when attaining Jupiter mass. Solid circles highlight the cases that match the meteoritic constraints. The accretion rate of solids is taken as that of SI08 in the case of orange circles, and of OG in the case of purple circles.}
\label{maps_a1}
\end{center}
\end{figure*}

The exact location at which Jupiter spent most of its formation is not known and is a subject of ongoing research \citep[e.g.,][]{Alibert05a, Walsh11}.
Recent work by \citet{Pirani19} suggests that Jupiter should have migrated inwards, from $a=9 -18$ au until its current position, to explain an asymmetry in the number of Jupiter's trojans. 
On the other hand, \citet{Desch18} claim that a much closer formation location is required (even inside the water iceline for some fraction of Jupiter's growth) to solve the CAI storage problem. 
Their model predicts the correct composition and formation location of 11 chondrite and 5 achondrite types. 

We start this section by testing two additional formation locations for Jupiter, the first inside the water iceline at $a= 1$ au; and the second beyond the current location of Jupiter, at $a= 10$ au.
  
Although our models do not include planetary migration, we have considered various parameters that affect the planetary growth such as formation locations, initial core masses and initial surface density of planetesimals. 
As a result, this study  covers a wide range of the parameter space that could emerge during migration. 
While planetary migration can affect the estimates for the accreted heavy-element mass, the migration history of Jupiter is poorly constrained. 
In addition, the processes and timescales associated with planetary migration are still being investigated and at the moment, choosing a specific prescription for it would make our results even more model dependent.
As a result, in this study we rather focus on {\it in situ} formation, which is complex enough, and hope to include planetary migration in future research.

We keep assuming large planetesimals (i.e., $\rp = 100$ km), but in the case of the formation inside the water iceline, we set the planetesimal mean density to be 3.2 g/cm$^3$, corresponding to a rocky composition \citep{Fortier13}. The results are summarized in Fig.\ref{maps_a1}. 
When the formation location is set inside the iceline, a clear difference with the growth at $a=$ 5 au is that the cases that fulfil the meteoritic time constraints are confined to an initial core mass of 16-17 \ME \, (see Appendix \ref{Mz_inside_iceline} for details). 
Also, when looking at all the simulations (not only those which fulfil the meteoritic constraints), we note that the amount of heavy elements accreted is not strongly dependent on the initial surface density of planetesimals, ranging between 13 \ME \, and  21 \ME \, when the mass of Jupiter is attained (bottom-left panel of Fig.\ref{maps_a1}), and reaching a maximum of 25 \ME\, when the initial core mass is $\Mcore$$_{,1}$=21 \ME.\,
This  means that a formation location inside the iceline is only consistent with the lower bound of the estimated heavy-element mass inferred from Jupiter's structure models, and therefore resembles a less likely formation scenario. 

For an assumed formation location of $a=10$ au, it is found that the total heavy-element mass depends strongly on the prescription for solid accretion. 
The prescription of SI08 gives a larger $\Mzdot$ for larger $a$ (see Eqs. \ref{vH_vscat} and \ref{vH_vdamp}).
In this case the amount of heavy elements is in the range of 20-80 M$_{\oplus}$, the upper value being beyond the estimated heavy-element mass of Jupiter for standard adiabatic models. 

\begin{figure*}
\begin{center}
	\includegraphics[width=\textwidth]{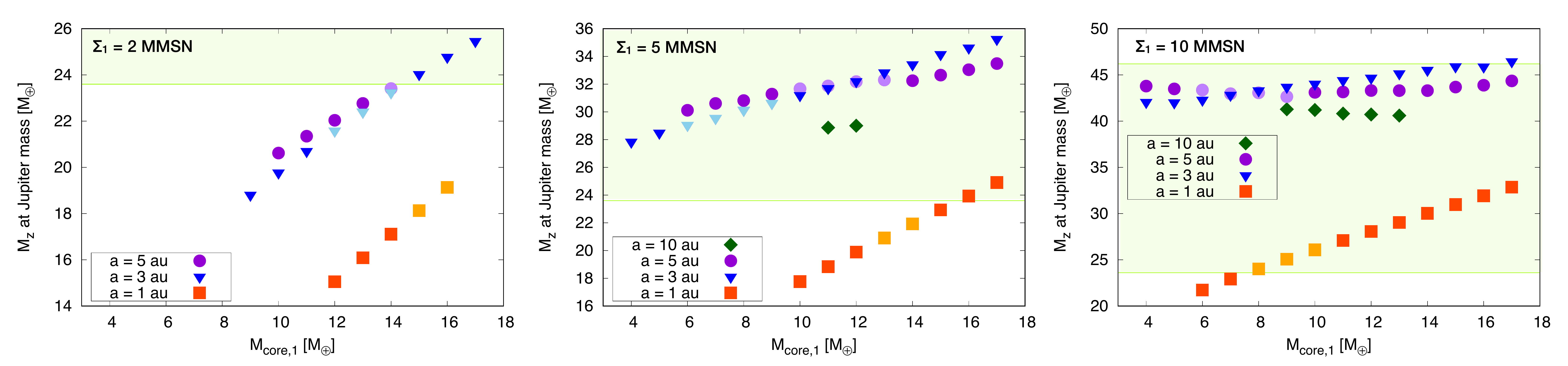} 
	\caption{Comparison of Jupiter's bulk metallic content between different formation locations, assuming the initial planetesimal surface density ($\Sigma_1$) corresponds to 2, 5 and 10 MMSN at their given location. 
	$\Sigma_1$ for each run is shown in Table \ref{table_Mz_MMSN}. The initial core mass for all cases is the same as the one assumed in the nominal setup: $4 \leq \Mcore$$_{,1} \leq 17 \, \Mearth$.
	The lighter colors remark the cases where the meteoritic constraints of K17 are met. The accretion of solids in the detached phase is given by SI08. }
\label{MMSN}
\end{center}
\end{figure*}

It should be noted that the reported accreted heavy-element masses correspond to the same assumed range of initial planetesimal surface density (5 $\lesssim \Sigma_1 \lesssim$ 25 \gcmcuad) for all the simulations. 
In classical disk models, where the solid surface density decreases outwards, the assumed $\Sigma_1$ might seem unfeasible for certain locations, or might give the impression that completely different scenarios are being compared. 
As a result, we plot, in Fig.\ref{MMSN}, the heavy-element mass attained at Jupiter mass for different formation locations, assuming the same $\Sigma_1$ in terms of the corresponding MMSN values at the given location. 
For this figure we have added a case of Jupiter forming at $a = 3$ au, as assumed by \citet{Desch18}. 

When $\Sigma_1$ is equivalent to an initial planetesimal surface density of only twice the MMSN (Fig.\ref{MMSN}, left panel ), at $a= 10$ au the growth is too slow and  Jupiter's mass cannot be reached within 10 Myr. 
At a= 1 au, the solid accretion rate is also low, so at Jupiter mass the heavy-element mass reaches a maximum of $\sim$20 $\Mearth$. The cases of $a=3$ and $a=5$ au reach the lower bound of the heavy-element content as inferred by structure models.

Also when $\Sigma_1$ is equivalent to an initial surface density of planetesimals of five times the MMSN (Fig.\ref{MMSN}, central panel), we confirm that a formation location of $a= 1$  au is very unlikely: 
only massive cores can reproduce the required enrichment of Jupiter, and for these cases the meteoritic time constraints of K17 are not met. Interestingly, we find that the case of $a= 10$ au is also very unlikely: only two initial core masses lead to the desired enrichment, and these cases also do not satisfy the meteoritic constraints. For $a= 3$ au and $a = 5$ au, a wide range of $\Mcore$$_{,1}$ can reproduce Jupiter's enrichment \citep[in agreement with][]{Shibata19}, and for both set of simulations, four different initial core masses match the meteoritic record. 

An intriguing aspect of all panels is the large differences between $a= 1$ au and $a=3$ au. For $a=1$ au, the longest phase of solid accretion happens after reaching the planetesimal isolation mass (see left panel of Fig. \ref{a1_a5}), while for $a \geq 3$ au, the planetesimal isolation mass is never reached, similarly to $a=5$ au \citep[also like in][]{Fortier07}. This makes all cases beyond $a=3$ au have a different planetesimal accretion rate, on average larger than at $a=1$ au, and decreasing with larger orbital distance.
The underlying reason for this behaviour is further discussed in Appendix \ref{Mz_inside_iceline} .

The right panel of Fig.~\ref{MMSN} shows results assuming $\Sigma_1$  = 10 $\Sigma_{\rm MMSN}$. In this extreme case in terms of available planetesimal mass, we see that a formation at $a = 1$ au would become possible. The same is true when a = 10 au, although without satisfying the meteoritic constraints. For a = 3 au all the assumed $\Mcore$$_{,1}$ reproduce Jupiter's enrichment, but no matching with K17 is found (Jupiter formation is too fast). For $a = 5$ au, both metallic and K17 constraints are still met.

Therefore, if we assume that Jupiter consists of 20-40 $\Mearth$ of heavy elements as suggested by standard structure models, it implies that after attaining the initial core ($\sim$5-15 \ME), Jupiter spent most of its time between $\sim$1-10 au (being 1 and 10 au unlikely locations).
Formation locations ranging between $a = 3$ and 5 au provide a very good match with Jupiter's current bulk metallicity and with the meteoritic constraints of K17; for a wide range of the assumed $\Sigma_1$, 
ranging from two to ten times the MMSN. For $\Sigma_1$ corresponding to one MMSN, the metallicity of the forming Jupiter is too low and the meteoritic constraints cannot be matched. 
This suggests that in our scenario, Jupiter's formation requires a "minimum planetesimal disk" corresponding to at least two times the MMSN.

It is however, hard to provide robust conclusions regarding Jupiter's formation location and growth history since the type of the accreted solids as well as their distribution within the disk (and its time dependence) are not well constrained. 

Finally, heavy elements can also be accreted by Jupiter after the gas disk has disappeared, and therefore our inferred metallicities should be taken as lower bound. 
Therefore it is important to acknowledge that in order to put tighter constraints on Jupiter's origin additional information is required. Further hints could come in the form of cosmos-chemical and isotopic data, comparative planetary science (i.e., comparing Jupiter and Saturn), formation models that account for the formation of the entire planetary system, and information from giant exoplanet data.

\subsection{Connection to Jupiter's internal structure}\label{Zetas}

While linking Jupiter's origin and its current-state interior is challenging, it is clearly desirable to compare the inferred heavy-element mass and its distribution from structure and formation models.
Recent structure models of Jupiter that match the gravitational moments measured by the \textit{Juno} spacecraft, 
suggest that Jupiter consists of  23.6 - 46.2 \ME\, of heavy elements, with a minimum core of 6.21 \ME~\citep{Wahl17}.
Interestingly, it was also found that the only structure models that fit the new gravitational data are ones with inhomogenous distribution of heavy elements. 
The core in these adiabatic models can be either compact (central region of $Z_{\rm core}$=1) or diluted ($Z_{\rm core}$ < 1). 

Alternatively, it is possible that the planetary interior is non-adiabatic and that the distribution of heavy elements changes gradually, with the heavy elements decreasing towards the atmosphere \citep[e.g.,][]{Leconte12, Vazan18, Debras19}.  
Indeed, such gradual Z-profile could be expected from formation models by core accretion when the vaporization of the incoming planetesimals is included \citep{Lozovsky17, Bodenheimer2018}. Note however that \citet{Lozovsky17} computed the heavy element mass deposition to pre-computed thermal/growth models, where the mass of all solids is added to the core. The work of  \citet{Bodenheimer2018} is the first to compute self-consistently the mass deposition with the change of composition in the structure equations. 
Their models, however, are only developed for the formation of planets inside the iceline accreting pure silicate-planetesimals. Moreover, they follow the planetary growth up to only $\sim$8 \ME. 
Thus, the role of compositional gradients in the giant planet formation and the expected distribution of heavy elements in young giant planets are still being investigated. 

\begin{figure}
\begin{center}
	\includegraphics[width=\columnwidth]{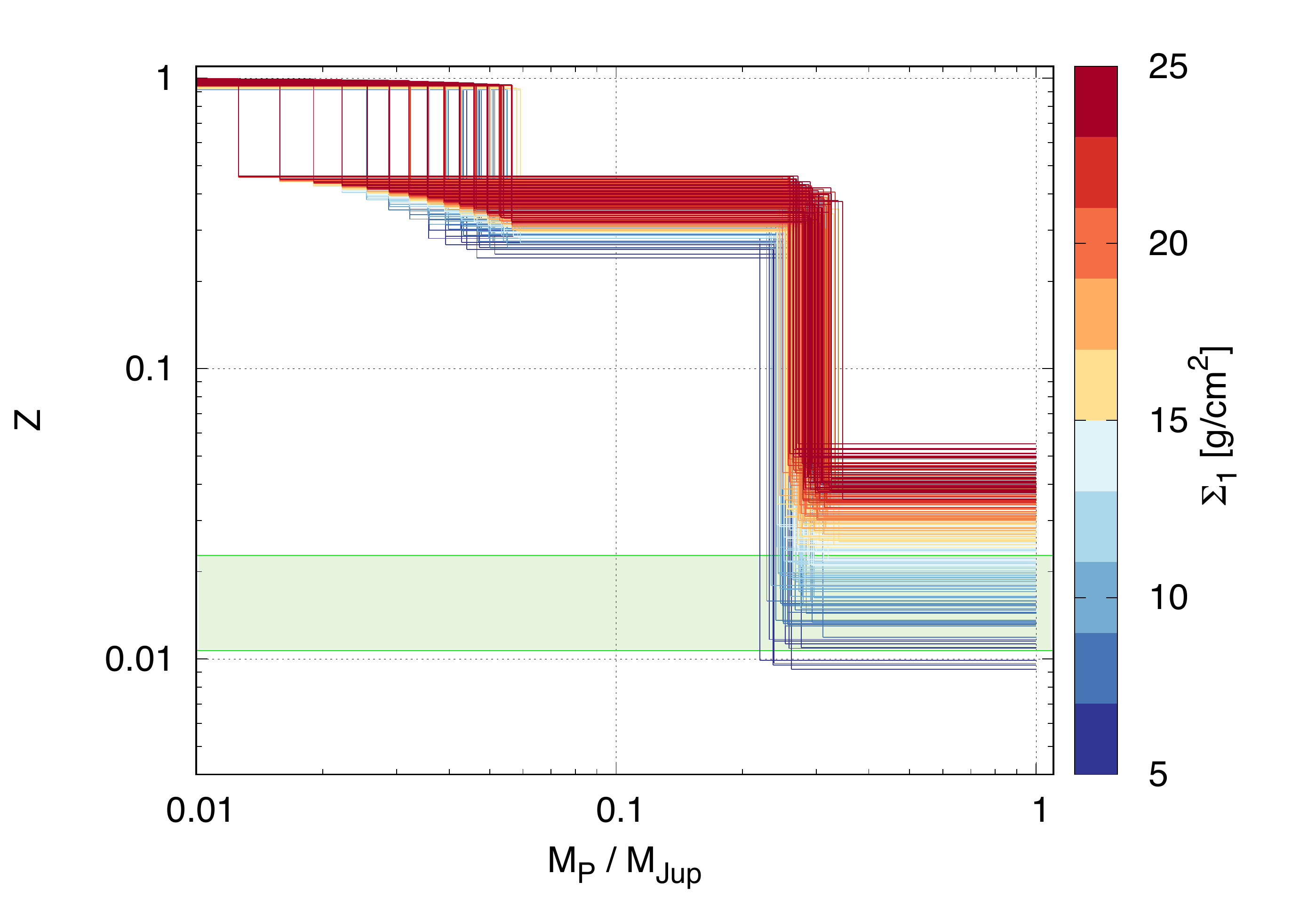}
	\caption{Jupiter's metallicity profile resulting from the 3 phases of accretion. All models attaining Jupiter mass within 10 Myr of disk evolution and corresponding to the nominal case with the SI08 prescription are shown.
	 The values of atmospheric metallicity compatible with the Galileo probe measurements are highlighted in green. }
\label{Zcum}
\end{center}
\end{figure}

In our formation models, we {\it assume} that the protoplanet's envelope is convective and it mixes the accreted heavy-elements quickly and efficiently. 
In all core accretion simulations, the first phase of growth is dominated by the accretion of heavy elements, resulting in a nearly pure-Z deep interior. 
Phase 2, of both heavy-element and gas accretion, creates an intermediate "layer" of lower metallicity, but still higher than the one expected during the third phase of runaway gas accretion, when most of Jupiter's mass is accreted, and the metallicity decreases. 
We illustrate this behaviour in Figure \ref{Zcum} where we show, for all the grid of simulations corresponding to the nominal case, the mean planetary metallicities resulting from the three mentioned phases of accretion. 
The core, presumably formed by pebbles, corresponds to the central region where the metallicity ranges between $0.91<Z_{1}\leqslant 1$. 
The middle layer, which is formed during Phase 2, has metallicity values of $0.24 \leqslant Z_{2} \leqslant$ 0.47. 
The outer layer, corresponding to Phase 3, when gas is accreted in a runaway fashion and planetesimal accretion continues, has a metallicity that ranges between $0.004 \leqslant Z_{3}\leqslant  0.01$ with the OG prescription and $0.009 \leqslant Z_{3} \leqslant 0.06$ with SI08 (see Fig.\ref{Zcum}).
The cases with the smallest planetesimal accretion during runaway (lowest $\Sigma_1$, see Fig.\ref{Zcum}) can lead to sub-solar atmospheric metallicities. 

A few of the formation paths we consider are consistent with the estimated metallicity of Jupiter's atmosphere. The atmospheric metallicity inferred from the Galileo probe is $Z_{\rm Gal} = 0.0167 \pm 0.006$  \citep[e.g.,][]{Debras19}. 
The nominal models that yield an outer metallicity in that range corresponds to all those with $ 6 \leqslant \Sigma_1 \leqslant  14$ \gcmcuad \, with the SI08 prescriptions . 
Note that these cases also lead to a total heavy-element content compatible with what is inferred from structure models that fit Juno data \citep[][see our Fig.\ref{heavies}]{Wahl17}. 
In particular, the models that fulfil the three constraints (Galileo measurement, structure models that fit Juno data, and K17) are the ones with a total heavy-element content of 24-33 \ME, of which 2-7 \ME \, are accreted during the detached phase (phase 3). 
  
\subsection{Mass-Metallicity relation: link to giant exoplanets}\label{MMrel}
\begin{figure}
\begin{center}
	\includegraphics[width=\columnwidth]{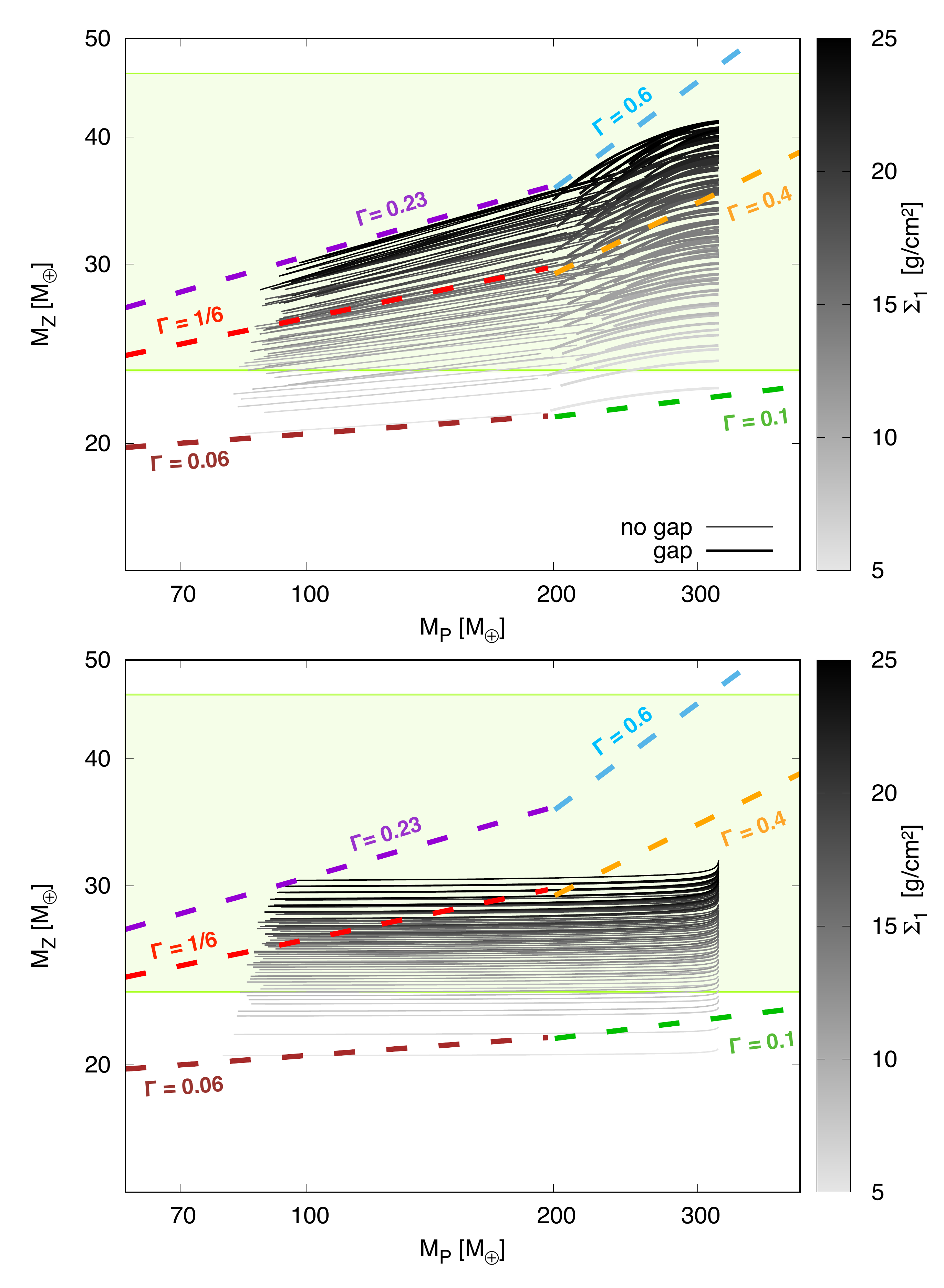} 
	\caption{Evolution of total amount of heavy elements as a function of planet mass during the detached phase, following SI08 (top) and OG (bottom). 
	For clarity, only cases that fulfil the meteoritic time constraints are shown.
	The final mass for each track is the mass of Jupiter.
	The grey color-bar indicates the initial planetesimal surface density (at time 1 Myr). 
	For the top panel, the thick(thin) lines show cases with(without) planetesimal gap.
	For both panels, the green shaded area delimits the range of heavy elements in Jupiter inferred by \textit{Juno}, and the color-dashed curves show different $M_Z = \mathcal{C}  \Mp^{\Gamma}$ relations, 
with ($\mathcal{C}, \Gamma$) = (15.5, 0.06) for the brown, ($\mathcal{C}, \Gamma$) = (12.33, 1/6) for the red, ($\mathcal{C}, \Gamma$) = (10.6, 0.23) for the purple,  ($\mathcal{C}, \Gamma$) = (12.5, 0.1) for the green, ($\mathcal{C}, \Gamma$) = (3.52, 0.4) for the orange, and ($\mathcal{C}, \Gamma$) = (1.48, 0.6) for the skyblue curve.} 
\label{MzMp_SI08}
\end{center}
\end{figure}

The metallic content of giant exoplanets was first studied by \citet{Guillot06}. 
Since then, it has been suggested that there is a correlation between the amount of heavy elements and the total mass of giant exoplanets \citep{MF11, T16}.  
The amount of heavy elements in a planet, however, cannot be directly measured and is inferred from models. The idea behind it is to take giant exoplanets with measured masses and radii, then together with the estimate of the stellar age one can run evolution models and search for the heavy-element mass (and therefore, bulk composition) that is required to fit the data.  
In particular, by computing the cooling of a sample of 47 giant exoplanets, \citet[][hereafter T16]{T16} found that the amount of heavy elements in giant exoplanets follows the power law:
$\Mz \sim \Mp^\Gamma$, with $\Gamma \approx 0.6$. Some studies have tried to reproduce such relation from formation models \citep{Mord14,Mord16}. 
In particular, \citet{Mord14} find from population synthesis that $\Gamma \approx 0.12-0.32$ (inferred from $\Gamma$ = 1 + $\alpha$  from the range of $\alpha$ shown in their Table 4). 
Recently, \citet{Hasegawa18} found from very simple analytical arguments that the T16 relation could be reproduced with planetesimal-gapped accretion. 
However, these simple estimates assumed a functional form of the SI08 prescription that is not the general one,  but for a uniform planet density, as clarified in \citet{Hasegawa18}, Sect. 5.1 (see Sect. \ref{SI08_methods}). 
We show in Appendix \ref{app_hasegawa} that the general expression of the gap-planetesimal accretion from SI08 leads to $\Gamma = 0.4$ when combined with gas accretion given by hydrodynamical considerations. 

In figure \ref{MzMp_SI08}, we show the $\Mz-\Mp$ relation we obtain from our simulations when adopting the SI08 prescription (top panel) and the OG prescription (bottom panel) during the detached phase.
For the top panel, we note that at the beginning of the detached phase, the planetesimal accretion occurs without opening a planetesimal gap, and afterwards, for planet masses larger than $\sim$150-200 \ME, a gap is formed.
This happens due to the decrease of the gas accretion rate originated by the opening of a gap in the gas disk \citep{Lissauer09}. The reduction of $\dot{M}_{\rm gas}$ abates the expansion of the feeding zone, which diminishes the amount of planetesimals that can access it. The planet mass at which the transition between non-gapped and gapped planetesimal accretion occurs depends on the gas accretion rate. Lower gas accretion rates in the detached phase make the transition occur at lower planet masses.

To visualize the range of possible $\Gamma$ for the $\Mz-\Mp$, the figure shows different power-law relations for the different regimes.
We note that the case of $\Gamma=0.6$ is extreme and can occur only for the gapped-planetesimal accretion in very metal-rich disks (large $\Sigma_1$), and for planets below Jupiter's mass.

Overall, we find a good agreement between our numerical and analytical results (the latter are shown in Appendix \ref{app_hasegawa}). For the gapped-planetesimal accretion, our numerical results yield $0.1 \lesssim \Gamma \lesssim$ 0.6, while our analytical estimate provides  
$\Gamma = 0.4$.  For the non-gapped planetesimal accretion, our numerical results follow a clear power-law with $0.06 \leq \Gamma \leq$ 0.23, while our analytical estime gives $\Gamma = 1/6$ (see Appendix \ref{app_hasegawa}). These slopes are shown in figure \ref{MzMp_SI08} for clarity.

Our results should not be interpreted as a disagreement with what is inferred from observations. 
Following one particular growth case (one curve in Fig.\ref{MzMp_SI08}) is not, strictly speaking, the correct way to compare with observations. 
The exoplanets in the T16 sample orbit stars with different properties, like different metallicities, masses and ages.
If, for instance, there were some bias towards more massive planets detected around more metallic stars (which is apparently the case, according to Fig.9 of T16), that would be a way to make the observed relation steeper according to our figure \ref{MzMp_SI08} (if a larger stellar metallicity implies a larger planetesimal surface density). 
In this regard, the SI08 prescription seems to be a promising recipe to reproduce the slope of the T16 mass-metallicity relation via population synthesis. This will be the topic of a follow-up work.

Another important aspect presented by T16 is the inferred absolute mass of heavy elements in the planetary sample. 
They find very large enrichments, with some exoplanets of Saturn mass reaching $\Mz\approx70\, \Mearth$, and of Jupiter mass reaching $\Mz \approx 100 \, \Mearth$.
From our formation models we find that this is very hard to reproduce. 

Only when assuming extremely high surface density of planetesimals we managed to reach $\Mz$ $\sim$ 80 \ME \, (Fig.\ref{maps_a1}, right panel). 
This case corresponds to assuming an initial planetesimal-to-gas ratio of $\sim 0.6$ at $a=10$ au, much higher than the usual $0.01$ value from classical models. 
This might suggest that giant planets with large metallic content could form at disk locations with extremely large planetesimal-to-gas ratios. 
Alternatively, large migration excursions could enhance the total amount of accreted solids \citep{ShibataEPSC}.

\section{Discussion} \label{sec_discussion}

\subsection{Formation Scenario: hybrid pebble-planetesimal model vs. pure pebble and pure planetesimal model}
Our model assumes the scenario proposed by K17 (but note that our results are weakly dependent on this assumption, Sect. \ref{results_Mz}), where it is claimed that carbonaceous and non-carbonaceous chondrites were separated not only in orbital distance but also during a lapse time of at least 2 Myr.
To fulfil this requirement, heat provided by planetesimal accretion is required, as shown in A18.

Note that even if the K17 interpretation were taken with skepticism, when trying to understand the growth of Jupiter with a pure pebble accretion model, one runs into the problem of explaining the metal content of Jupiter inferred by Juno. This should be $\sim$24-46 \ME.
In a pure pebble accretion model, when the protoplanet reaches the pebble isolation mass, the accretion of heavy elements is halted, because the pebbles are trapped in the pressure bump generated in the vicinity of the protoplanet.  This means that the amount of heavy elements accreted by Jupiter should correspond  to the pebble isolation mass. 
The pebble isolation mass depends on the disk aspect ratio, viscosity, and radial pressure gradient \citep{Sareh18, Bitsch18}. A pebble isolation mass larger than $\sim$25 \ME \, (approximate minimum amount of $\Mz$ inferred by Juno) requires disks with very large aspect ratios ($h/H > 0.05$) and/or very large viscosities ($\alpha>10^{-3}$) \citep{Sareh18, Bitsch18}. 
Thus, a pure pebble accretion model for Jupiter seems to run into difficulties when trying to explain the actual metal content of the planet.

It should be noted that pure planetesimal accretion models have tackled the problem of forming Jupiter since decades \citep{P96, Alibert05b, Hubi05, Lissauer09, Fortier09, Guilera11}. 
Despite of the long core formation timescales associated with this scenario, especially when the gravitational interaction between the embryo and planetesimals is considered \citep[oligarchic growth,][]{Inaba01,Fortier07}; these models, under certain assumptions, can lead to the formation of Jupiter. 
Typically, it  is required to assume small planetesimals in order to reach crossover mass before disk dissipation \citep{Fortier13, Guilera11}. 
Shortening the formation timescale can be achieved when considering additional physical processes such as opacity reduction due to grain growth and settling \citep[e.g.,][]{Movshovitz10}, or heavy-element pollution of the planetary envelope \citep{HI11, Venturini16}. 
The amount of heavy elements in Jupiter predicted by these different planetesimal-based models has a large range. For instance, models that invoke low dust opacities can form a Jupiter with a core of  $\sim$5 $\Mearth$ \citep{Hubi05}, models that assume large dust opacities and large surface densities of solids can yield a core of up to $\sim$40 \ME \citep{Guilera10}. Still, it s important to remark that no pure-planetesimal accretion model can form Jupiter with 100 km planetesimals when the interaction between embryo-planetesimals is correctly accounted for \citep{Fortier13, Guilera14}. This is overcome in the hybrid pebble-planetesimal accretion model.
 
 Finally, it should also be noted that several studies suggested that the standard Phase 2 of giant planet formation, also in the planetesimal accretion scenario, can be extremely short or even non-existing \citep{Zhou2007,SI08}. 
 In that case, when the formation of a gap in the planetesimal disk is considered, the core mass is expected to be relatively small with a significant planetesimal accretion when the planet reaches a mass of $\sim$ 100 $M_{\oplus}$ \citep{SI08}. 
 While this is an interesting option, we suggest that further calculations which account for the planetary internal structure and limit the planetesimal supply must be considered. In addition, it is yet to be shown that such a formation scenario with no Phase 2 is consistent with the cosmo-chemical  constraints discussed in this study.

\subsection{Water Abundance and formation location}
The \textit{Juno} spacecraft is also expected to measure the water abundance in Jupiter using the microwave radiometer (MWR) which probes Jupiter's deep atmosphere at radio wavelengths \citep{Bolton17}.
Data from MWR is still being collected, and constraints on the water abundance are expected in the near future. It should be noted, however, that 
Juno's measurements are limited to the uppermost 100--200 bars, and might not represent the bulk (e.g., Helled \& Lunine, 2014). 

In order to infer the global water abundance in Jupiter, one must assume that the heavy elements are homogeneous within the planet, an assumption that has recently been 
challenged by Jupiter's structure models fitting Jupiter's gravitational field \citep{Wahl17, Vazan18, Debras19}. 
In this case the MWR estimate is likely to provide a lower bound for the water abundance. 
 
In Sect.\ref{sec_a1} we analyzed the possibility of Jupiter spending most of its formation time inside the water iceline. 
We found that in such case the heavy-element mass is in the lower bound of the range inferred from structure models. 
In terms of pebble isolation mass, at 1 au only a value of $\sim$ 6 \ME \, is expected \citep{Lambrechts14}. 
This means that to fulfil the K17 scenario, a core of 16 \ME \, should have migrated inwards, from beyond the iceline. 
In that case, half of it should be made of ices \citep[e.g.,][]{Amaury15}, which would give Jupiter a maximum amount of $\sim$8 \ME \, of water, and probably very concentrated towards the center of the planet (the accretion of  dry solids in our simulations would be by planetesimals, in Phase 2 and Phase 3 of growth). 
Note, however, that this limit would increase if Jupiter accreted, post-disk dispersal, icy planetesimals/embryos.

From the simulations beyond the iceline (Figs.\ref{maps_nominal},\ref{MzMp_SI08}), we find that Jupiter could have a range of $\sim20-75 \, \Mearth$ of heavy elements. 
The upper value is linked to the formation test case at $a=10$  au (Sect.\ref{sec_a1}). 
In the nominal formation location of 5 au, the upper limit on Jupiter's heavy elements is found to be $\Mz = 45 \, \Mearth$ (Fig.\ref{heavies}).
According to condensation sequence studies, half of these heavy-elements could be volatiles \citep{Amaury15}, which, if we neglect the presence of other volatiles than water, sets a maximum amount of water in the range of $\sim$ 12-23 \ME. 
If beyond the iceline water resembles at least 30\% in mass of the condensed material, 
 this leads to a minimum amount of water of  $\sim$ 8 \ME \ for Jupiter. 

The above considerations imply that if the maximum amount of water in Jupiter is below 8 \ME, this would suggest that Jupiter spent a few million years within the iceline.  
In case the water content were confirmed to be above 10 \ME, the conclusions are more blurry, because Jupiter could have spent some time within the iceline, but still have been polluted by ices after the disk dispersal.
Still, if Jupiter had spent phase 2 and 3 inside the water iceline, it would have a larger rock-to-ice ratio compared to a full formation beyond the iceline. In the latter, the rock-to-ice ratio should be close to 1. In the former (phase 2 and 3 within the iceline), Jupiter would have started with 8 \ME \, of rocks and the same amount of ices, then accreted maximum 8 \ME \, of rocks (to get a total of 24 \ME \, of heavy elements, as expected by Juno). 
Thus, the rock-to-ice ratio would be closer to  $\sim$2.

Note that the above discussion assumes  the water iceline to be the location in the disk where planetesimals form with temperatures of $\sim$170 K, and that the planetesimals do not change their composition during the disk evolution. In reality, the water snow line is expected to move as the disk evolves \citep[e.g.,][]{Oka11}, but this would modify mainly the composition of pebbles \citep{Morby16}. In addition, the water abundance in the accreted solids is unknown. Therefore, linking Jupiter's water abundance with its formation process is non-trivial: it depends on the formation location of the planet, its migration history, and the composition of the accreted planetesimals/pebbles. As a result, the inferred numbers should be taken as a guideline as they are derived under simplified assumptions.  
Finally, if certain evaporative mechanism operates efficiently, like heating from Al-26 \citep{Lichtenberg19}, the above inferences would be only an upper bound on water content. Jupiter could be much drier than expected even if it spent phase 2 and 3 beyond the water iceline. In summary, constraining the planetary origin from water abundance estimates is a difficult task, and it suffers from various uncertainties and degeneracies in theoretical calculations.

Further constraints might come from linking cosmochemical data with astrophysical models. Indeed, \citet{Desch18} show that in order to solve the CAI storage problem, the second phase of Jupiter's growth had to be spent between $a=$2.9 and $a=$3.2 au. 
If Jupiter had been farther out, the amount of CAIs to form chondritic chondrites beyond Jupiter would have been too scarce. 
We found formation models of Jupiter that match both the estimated metallicity of the planet and the K17 constraints for $1<a<10$ au, being $a$=3 au a perfectly possible location for Jupiter to spend phase 2 (see Sect.\ref{sec_a1}). 
Many existing models explain different aspects of Jupiter's observed properties, but clearly more work on linking meteoritic data, disk thermal evolution, simultaneous formation of Jupiter and Saturn with the correct metallic content, and special migration regimes \citep[e.g,][]{Masset01} is required  in order to develop a coherent picture of the history of our Solar Sytem.

\section{Conclusions}
We simulate Jupiter's growth and followed the heavy-element accretion for various model assumptions. 
Many of our models can reproduce the amount of heavy elements in Jupiter as inferred from structure models \citep{Wahl17}.  

Similarly to \citet{A18}, we show that the initial phase of core formation (phase 1) is likely to be dominated by pebble accretion, followed by a second stage (phase 2) of planetesimal accretion. 
By modelling the second stage coupled self-consistently with gas accretion up to Jupiter mass (phases 2 and 3 in our scenario), we find that the meteoritic time constraints derived by K17 can be successfully matched with the accretion of 100 km-size planetesimals. This is an important finding, because it is the first time that large planetesimals are shown to play a crucial role in Jupiter's growth and enrichment when the excitation of planetesimals by the embryo is consistently accounted for (oligarchic growth regime).
It is also important to note that the total heavy-element mass in Jupiter inferred from our simulations is independent on the assumption that Jupiter is responsible for reconnecting the carbonaceous and non-carbonaceous reservoirs (K17, A18). 

We suggest that a formation location inside the water iceline after reaching pebble isolation mass is less likely for Jupiter, since the amount of heavy elements that can be accreted in the inner regions of the disk is much lower than at 3-5 au, 
making the total heavy-element mass accreted by Jupiter in the lower-bound of the heavy-element mass estimates from structure models. 
Our calculations favor Jupiter's formation at $1< a <10$ au after pebble isolation mass is reached. 

We find that our hybrid pebble-planetesimal model can produce a Jupiter that matches the estimates for the planet's bulk metallicity and the meteoritic time constraints derived by K17. 
In addition, when examining the average metallicity for a given formation phase, and assuming that the layer corresponding to this formation phase is homogeneously mixed, we can reproduce Jupiter's atmospheric metallicity as inferred by the Galileo probe.
For these models, the heavy element-mass accreted  during the detached phase is  found to be $M_Z\approx 2-7$ \ME. 

We also find that Jupiter can accrete $\sim$1-15 $\Mearth$ of heavy elements during its final formation phase.
The exact value depends on the initial surface density of planetesimals and their dynamical properties, which are highly unconstrained. 
We find that if the initial planetesimal surface density corresponds to approximately  two times the MMSN, the accreted heavy-element mass during Phase 3 would be 
 $\sim$1 \ME, and of $\sim$6-15 \ME\, for ten times the MMSN. 
This is an important finding since it has direct implications of our understanding of the enrichment of giant planets in general. If a non-negligible amount of heavy elements can be accreted in this last stage of formation, it would lead to a higher metallicity in the outer planetary envelope. For the case of Jupiter, we find that such "late enrichment" can yield an envelope metallicity of $\sim$0.5 to 3 times solar. This results suggest that the diversity in planetary composition is a natural outcome of planet formation models, with the exact composition depending on the exact formation location and growth history.

We have, additionally, investigated the link to the metallicity of giant exoplanets. We find that the very high bulk metallicities inferred for many exoplanets in the \citet{T16} sample are hard to reproduce. 
It is found that an extremely high initial planetesimal surface density, corresponding to an initial planetesimal-to-gas ratio of $\sim$0.6, could yield $\Mz \sim$ 80 \ME \, at Jupiter mass and $a=10$ au. 
This could suggest a migration history for the metal-rich warm giant exoplanets, as recently proposed by \citet{ShibataEPSC}.

Finally, we examined the slope of the Mass-Metallicity relation of giant exoplanets, and found analytically that  $\Mz \sim \Mp^\Gamma$, with $\Gamma \approx 1/6$ when no gap in the planetesimal disk is created, and with $\Gamma \approx 0.4$ with planetesimal gap. These values are in general agreement with our numerical results, but are 
lower than the 0.6 value found by T16. The solid accretion rate provided by \citet{SI08} could lead to a better match with T16 when implemented in population synthesis models, and we hope to address this point in future work.

\bigskip

\textit{Acknowledgements.}
We thank Sho Shibata and Masahiro Ikoma for enlightening discussions, and Octavio Guilera for relevant suggestions.
We also thank the anonymous referee for the insightful comments, which helped us to improve the manuscript.
This work has been carried out, in part, within the framework of the National Centre for Competence in Research PlanetS, supported by the Swiss National Foundation. 
R.H. acknowledges support from SNSF grant 200021\_169054.


\begin{appendix}

\section{Details on planetesimal accretion: the importance of the capture radius and the initial large core mass}\label{app_Rcap}
A long standing problem in planet formation has been the slow planetary growth when the solid accretion is given solely by large planetesimals. 
Indeed, simulations show that the formation timescale of gas giants typically exceeds disk lifetimes \citep{P96, Fortier07}. 
Two broad solutions for this  "timescale problem" were proposed: 
\begin{enumerate}
\item[(i)] Planets grow mainly by accretion of small planetesimals (100 m- 1 km) in disks with relatively large metallicities \citep{Fortier09, Guilera11, Fortier13, Alibert13}
\item[(ii)] Planets grow predominantly by cm-size pebbles \citep{Lambrechts14}. 
\end{enumerate}

One problem with the first scenario is that planetesimals seem to be formed large ($\sim$100 km) by streaming instability \citep[e.g, ][]{Simon17}. 
In addition, when collisions are included, planetesimals in the size range of $\sim$ km do not live long \citep{Kobayashi10}.
On the other hand, the pure pebble accretion scenario runs into other type of difficulties. First, pebbles drift fast towards the central star, and protoplanetary disks have limited sizes of $\sim$10-200 au \citep{Andrews10}, so the prevalence of pebbles in the disk during the $\sim$2-10 Myr of disks's lifetime is not guaranteed. 
Second, as the disk evolves and cools, icy pebbles can reach the inner regions of the system, polluting it with volatiles \citep{Morby16}. 
Recent interpretation of the Kepler data suggests that super-Earths are dry \citep{Owen17}, which is at odds with a pure pebble accretion scenario.
In addition, pebble accretion is either too efficient (disks with low viscosity) or too inefficient (disks with high viscosity), which leads to the formation of either too many gas giants or too many embryos \citep{Lin18, Venturini17}.
Indeed, intermediate-mass planets (Neptunes), which are very abundant \citep{Batalha13, Suzuki18}, are hard to form only with pebbles \citep{Venturini17}.

A hybrid pebble-planetesimal accretion scenario for giant planet formation does not only provide a solution for the isotopic anomalies of carbonaceous and non-carbonaceous meteorites \citep{A18}, 
but it also has the potential to overcome the aforementioned problems. 
An initial dominant phase of pebble accretion is plausible, given that disks are expected to be very abundant in pebbles at the beginning of their lifetimes \citep[e.g.,][]{LJ14}.

In the light of this discussion, it is important to clarify why in the new pebble-planetesimal accretion scenario, the accretion of large planetesimals is consistent with the formation of a gas giant. 
The key for 100 km- planetesimal accretion to be successful is twofold: 
\begin{enumerate}
\item The initial core mass cannot be the usual lunar mass \citep[value that came from the original pure N-body simulations of oligarchic growth,][]{KI98}. It has to be at least 4 $\Mearth$. This core can be formed easily by pebble accretion. 
\item  The atmospheric enhancement factor for the accretion rate (the capture radius) has to be considered \citep{InabaIkoma03, Fortier09}. 
\end{enumerate}

\begin{figure}
\begin{center}
	\includegraphics[width= 0.95\columnwidth]{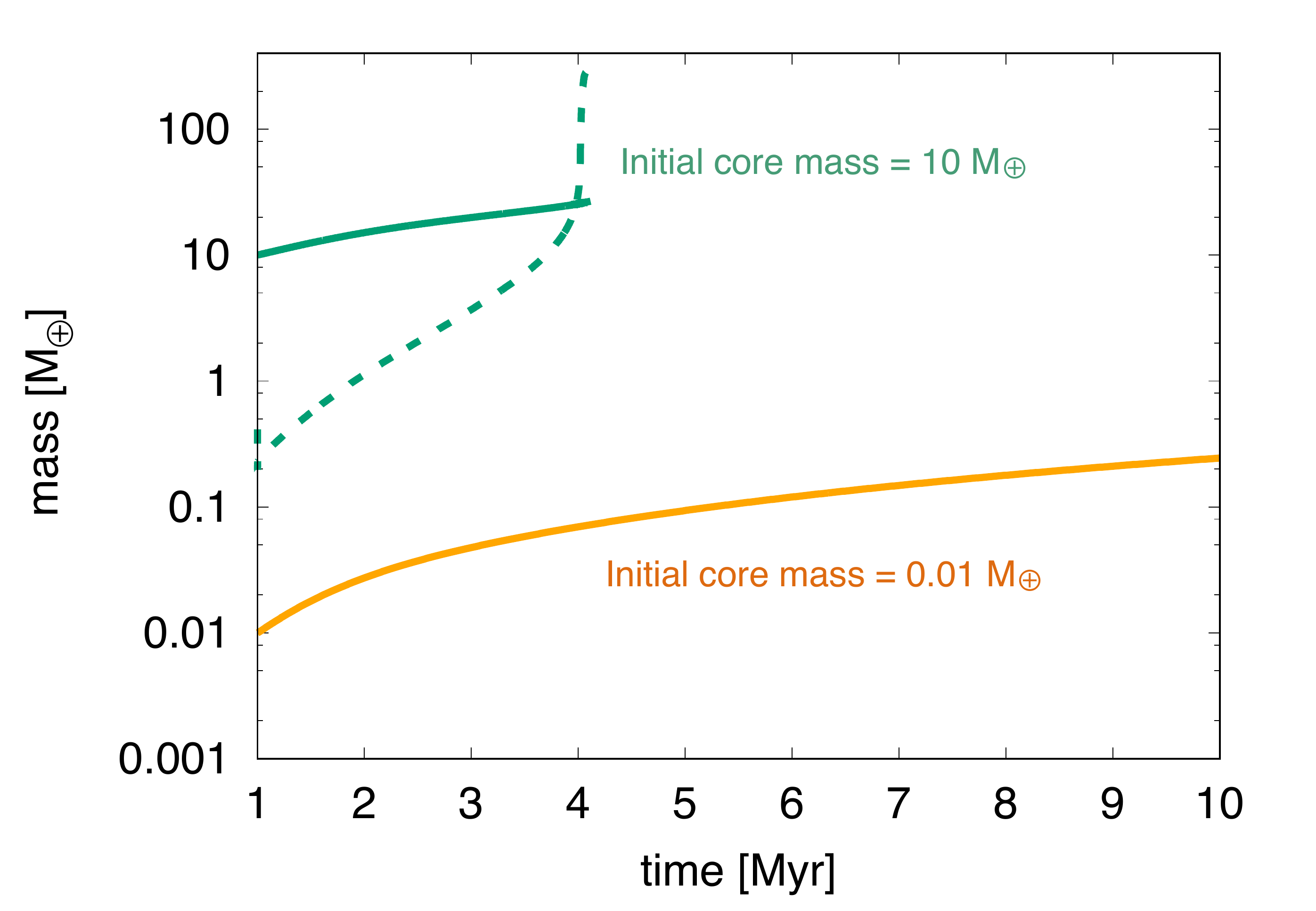} 
	\caption{ Growth of a planet corresponding to the nominal setup, with $\Sigma_1$ = 15 \gcmcuad. 
	The green curves assume an initial core mass of  $\Mcore$$_{,1}$= 10 \ME; and the orange curves an initial core mass of $\Mcore$$_{,1}$= 0.01 \ME.
	Solid lines indicate the growth of the core, dashed lines the growth of the envelope. The envelope mass of the latter case is always smaller than 0.001 \ME. }
\label{moon}
\end{center}
\end{figure}

\begin{figure}
\begin{center}
	\includegraphics[width= \columnwidth]{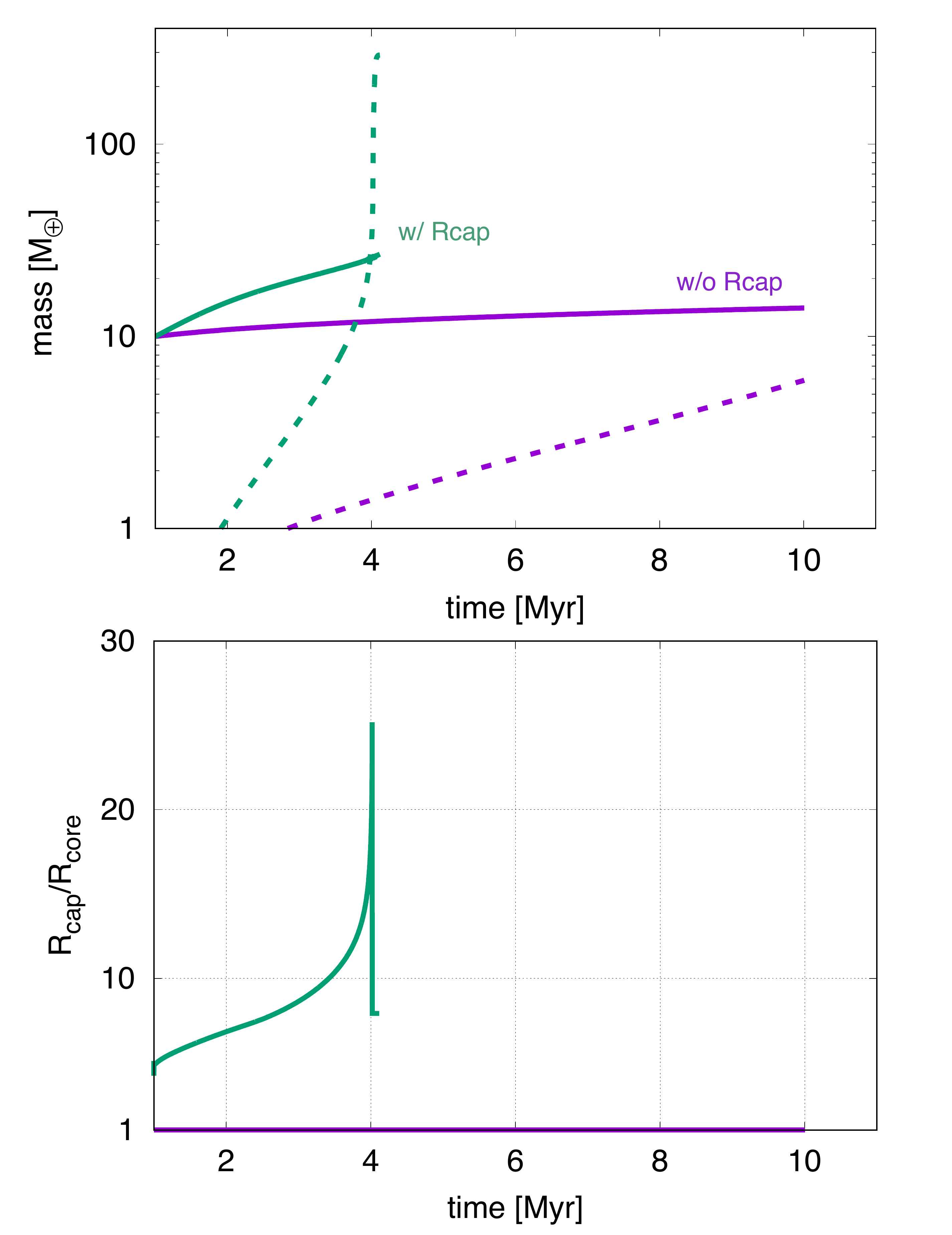} 
	\caption{
	{\it Upper panel:}  Growth of a planet for  $\Mcore$$_{,1}$= 10 \ME \,and  $\Sigma_1$ = 15 \gcmcuad, with (green) and without (purple) the effect of the capture radius. 
	The solid lines show the growth of the core and the dashed ones the growth of the envelope.  
	{\it Bottom panel:} The evolution of capture radius in units of core radius for the growth cases shown in the top panel. In the purple case the capture radius is always fixed to the core radius. 
	The green curve drops to a radius of 1.8 R$_{\rm Jup}$ when the detached phase starts.}
\label{Rcap_fig}
\end{center}
\end{figure}

We illustrate the importance of these two effects in Figures \ref{moon}. and \ref{Rcap_fig}. 
The former shows the planetary growth with the nominal set up, and initial planetesimal surface density of $\Sigma_1$ = 15 \gcmcuad. One growth-case assumes an initial core mass of $\Mcore$$_{,1}$= 10 $\Mearth$ and the other, of $\Mcore$$_{,1}$= 0.01 \ME. 
When the core starts its growth with a lunar mass, at $t = 10$ Myr it reaches a core mass of only $\Mcore = 0.24 \, \Mearth$, whereas in the case where $\Mcore$$_{,1}$= 10 \ME, crossover mass (25 $\Mearth$ in this case) is attained at $t = 4$ Myr. 
This means that for the case that starts with a lunar embryo, the average accretion rate of solids is 0.23 $\Mearth$ per 10 Myr (or $2.3\times10^{-8}$ \ME/yr); 
while in the nominal case the mean accretion rate of solids is 15 $\Mearth$ per 4 Myr (or $3.75\times10^{-6}$ \ME/yr ). The planetesimal accretion rate during the OG goes with R$_H^2 \sim \Mp^{2/3}$, which gives a ratio between the two starting embryos of $(10/0.01)^{2/3}$ = 100. This is practically the same difference in ratio that we obtain between the two mean planetesimal accretion rates.

The effect of the capture radius on the core's growth is demonstrated on Fig.\ref{Rcap_fig}, where, as before, $\Sigma_1$ = 15 \gcmcuad and $\Mcore$$_{,1}$=10 \ME. 
In one simulation the capture radius is computed and included in the cross section of accretion, and in the other case, the capture radius is set to the core radius for the whole simulation (i.e, the enhancement factor of accretion is neglected).
 The effect on the core growth is shown on the top panel of Fig.\ref{Rcap_fig}. The bottom panel shows the value of the capture radius in units of core radius for both setups. 
 We can see that without accounting for the capture radius, the core grows too slowly to attain crossover mass before the dissipation of the disk.

\section{Difference in solid accretion between growing Jupiter inside or beyond the iceline}\label{Mz_inside_iceline} 
\begin{figure*}
\begin{center}
	\includegraphics[width=0.8\textwidth]{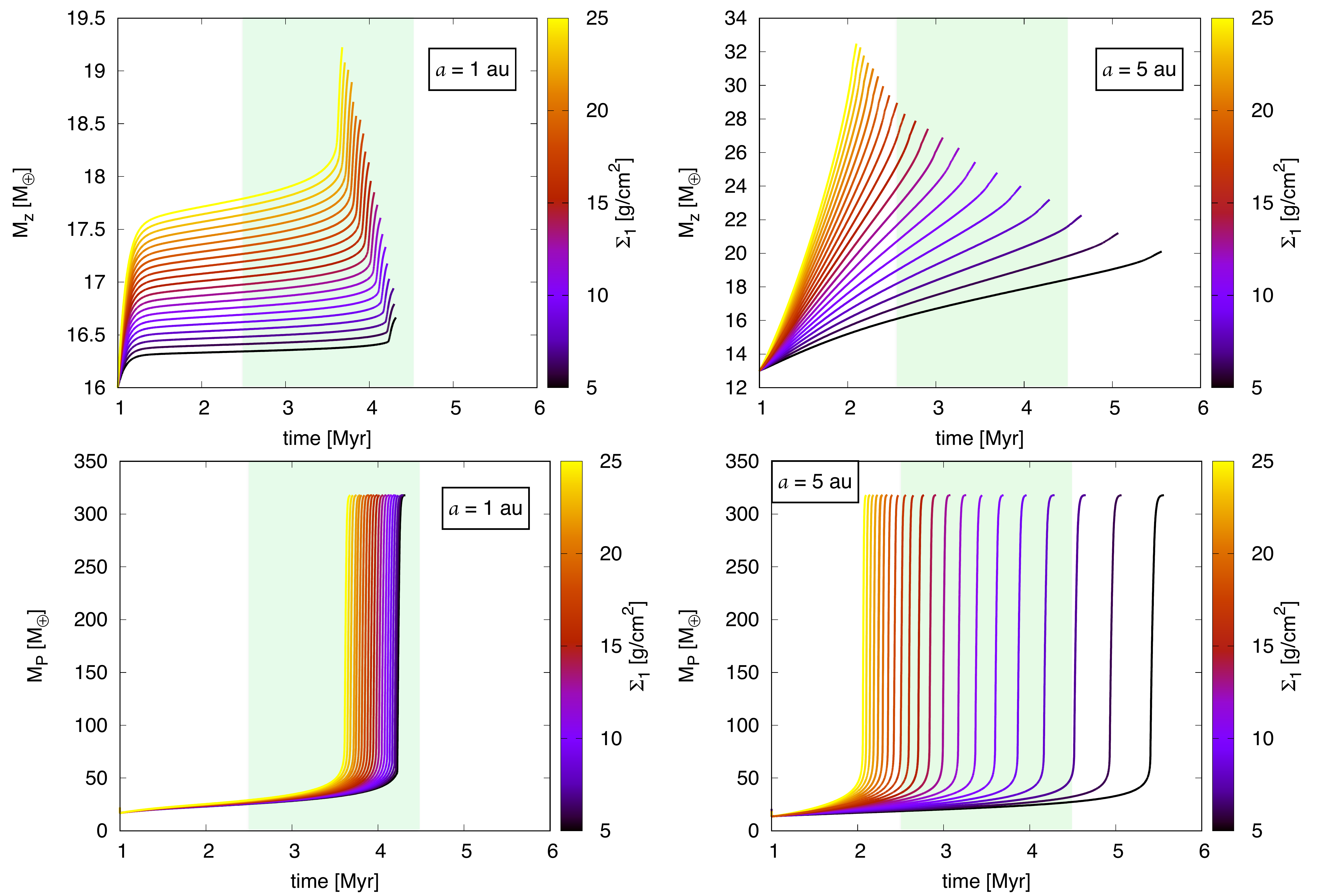}
	\caption{Growth of planet at $a$= 1 au (left) and $a$= 5 au (right). For clarity, only cases with an initial core of 16 \ME\, and 13 \ME \, are shown (left and right panels, respectively). 
	Top panels: evolution of the mass of heavy elements. Bottom panels: evolution of the total planet mass. Color-bar: initial surface density of solids. 
	Green shaded area: region where the meteoritic time constraints are met (up to $\Mp\sim50 \, \Mearth$). }
\label{a1_a5}
\end{center}
\end{figure*}

We mentioned in Sect \ref{sec_a1} that the formation scenario at $a = 1$ au is unfavourable for Jupiter's formation due to the relatively low bulk metallicity of Jupiter.  
Although this formation scenario is not very likely, it is interesting to understand the different behaviours between Figs. \ref{maps_nominal} and \ref{maps_a1}. 
We therefore compare the growth of the heavy elements and total planet mass for the cases of 1 au and 5 au in Fig.\ref{a1_a5}.
For clarity, only cases where $\Mcore$$_{,1}$ = 16 \ME \, ($a= 1$ au)  and $\Mcore$$_{,1}$  = 13 \ME \, ($a = 5$ au) are shown.
The different influence of $\Sigma_1$ at both locations is very clear. The green-shaded area shows the cases that match the meteoritic time constraint. 
For an equal  $\Mcore$$_{,1}$, all the values of $\Sigma_1$ produce growth paths that match the meteoritic time constraints for $a$=1 au, whereas for  $a$ = 5 au, this depends on the value of $\Sigma_1$. 

The growth of the heavy-element mass is very different in both cases. In the former, the core grows quickly at the beginning, consuming all the planetesimals available in its feeding zone in the first $\sim$300 thousand years.
The planetesimal isolation mass is a strong function of the semi-major axis, with a milder dependance on the initial planetesimal surface density  \citep[$\Miso \sim a^3 \Sigma_1^{3/2}$, ][]{P96}. 
This implies that at $a$=5 au, the planetesimal isolation mass is $\sim$ 100 larger than at $a$=1 au ($\Sigma_1$ being constant). Consequently, the case at $a$=5 au never reaches the planetesimal isolation mass: there are always enough planetesimals to be accreted in the feeding zone. Indeed, we checked that at a= 5 au, the amount of planetesimals accreted at each time step until crossover mass never exceeds 1\% of the total amount of planetesimals available in the feeding zone at that time.
The duration of phase 2\footnote{The phases in this Appendix refer always to the classical phases of \citet{P96}, which differ from the 3 phases introduced in the main text (Sects. \ref{intro} and \ref{method}).} is less dependent on $\Sigma_1$ than phase 1. The main factor influencing the duration of phase 2 is the composition of the envelope, which is the same for all the cases plotted here.
It is exactly the fact that the planetesimal isolation mass is so much smaller at $a$= 1 au than at $a= 5$ au what makes it difficult for a giant planet to accrete substantial amounts of solids in close-in orbits: there is simply not enough material.

\section{An Analytical estimate of the $\Mp$-$\Mz$ relation from formation}\label{app_hasegawa} 
In this section we derive analytically the  $\Mp$-$\Mz$ relation assuming the accretion rate of solids is given by \citet{SI08} (SI08) and that the gas accretion is dictated by the supply of the disk \citep{Tanigawa02}. 
We follow the same approach as \citet[][hereafter H18]{Hasegawa18}, but using the general form of the planetesimal accretion from SI08 (Eqs. \ref{nogap} and \ref{gap}), 
and assuming that the protoplanet's radius has shrunk to a fixed value, a result that is found by several studies during the detached phase \citep{Lissauer09, Mordasini13}.
It should be noted that H18 assume that the density of the protoplanet remains uniform during this final stage of accretion, i.e, $\Rp \sim \Mp^{1/3}$, which is not what is found when computing the actual radius during the detached phase.

We start by analysing the gapped-planetesimal case.
From Eq.\ref{gap} we have the following functional dependance,
\begin{equation}\label{gap_2}
	\MzdotG \sim {\rho}^{1/2} \, \Rp^2 \, \bigg(\frac{v_{\rm H}}{v_{\rm damp}}\bigg)^{\alpha}.
\end{equation} 
 
Considering that $\tau_g \sim \Mp^D$ and using Eq.\ref{vH_vdamp} only with the mass dependence (as in H18), the mass dependence of Eq.\ref{gap_2} reads,
\begin{equation}
 \MzdotG \sim \Mp^{1/2 - \alpha(1/6 + D)} 
\end{equation}  

The metal content ($\Mz$) as function of planet mass ($\Mp$) can be obtained from integrating d$\Mz$/d$\Mp$, which is given by:
\begin{equation} \label{derivMzMp}
\frac{d\Mz}{d\Mp} = \Mzdot \frac{dt}{d\Mp} \sim  \Mzdot \frac{\tau_g}{\Mp}
\end{equation}

Using the above expressions, we obtain, for gapped-planetesimal accretion, d$\Mz$/d$\Mp$ = $\Mp^{\gamma}$, with 
\begin{equation}
\gamma = -\frac{1}{2} - \frac{\alpha}{6} + D(1-\alpha)
\end{equation}

Hence, when integrating d$\Mz$/d$\Mp$ in the planet mass, we get $\Mz = \Mp^{\Gamma}$, where $\Gamma = 1 + \gamma$, i.e: 

\begin{equation}
\Gamma = \frac{1}{2} - \frac{\alpha}{6} + D(1-\alpha). 
\end{equation}

Plugging-in the values for gap-planetesimal accretion ($\alpha$ = 1.4) and for the timescale to accrete gas  given by the supply of the disk (see H18) ($D = -1/3$), 
we obtain  $\Gamma = 0.4$. Note that this exponent of the  $\Mp$-$\Mz$ is not far from the one derived by T16 ($\Gamma = 0.61 \pm 0.08$) and is actually more consistent with our numerical results (top panel of Fig.\ref{MzMp_SI08}).

Analogously, for the non-gapped planetesimal case, the functional dependence of the accretion rate of solids goes as (Eq.\ref{nogap}):
\begin{equation}\label{nogap_2}
	\MzdotNG \sim {\rho}^{1/2} \, \Rp^2 \, \bigg(\frac{v_{\rm H}}{v_{\rm scat}}\bigg)^{\alpha}.
\end{equation}
where $\alpha$ =0.8 

Applying again Eq.\ref{derivMzMp} and assuming the accretion rate of gas is limited by the disk supply, we reach, by using Eqs.\ref{nogap} and \ref{vH_vscat}:
\begin{equation}
\Gamma = \frac{1}{2} - \frac{\alpha}{3} + D(1-\alpha). 
\end{equation}
Plugging-in the corresponding values of $\alpha = 0.8$ and D=-1/3, we obtain $\Gamma = 1/6$.

\section{Heavy element accretion in each phase of Jupiter's growth} \label{app_summary_all_phases}
Table \ref{table_Mz} lists that heavy-element accretion mass during each phase of Jupiter's growth as inferred from all the simulations presented in this study.
The table is constructed as follows. 
For each run, the maximum and minimum ranges of the assumed $\Sigma_1$ are shown, and for each $\Sigma_1$, the maximum and minimum values of $\Mcore$$_{,1}$  that yield to a Jupiter mass planet within 10 Myr of disk evolution. 
Thus, for each run, typically four extreme cases are shown, and all the corresponding intermediate values of ($\Sigma_1$, $\Mcore$$_{,1}$) yield an accreted amount of solids that lie in between the shown boundaries. 
For one case there is only one value of $\Mcore$$_{,1}$ for a given $\Sigma_1$. For that case, a lower $\Mcore$$_{,1}$ resulted in a planet that has not reached a Jupiter mass within 10 Myr, and a larger value of $\Mcore$$_{,1}$ yielded a supercritical planet.  
By comparing the amount of solids accreted in the different phases, one can see that the ones accreted during the attached Phase 3  have always very low values. This is because the detached phase starts typically soon after the crossover time. For a few cases in the run of envelope enrichment, the crossover mass is reached just after the detached phase starts, meaning that there is no \textit{attached Phase 3}.

Note that for a given $\Sigma_1$, typically the amount of heavy elements accreted during Phase 2 is larger for the smaller $\Mcore$$_{,1}$.
 This is because the time duration of Phase 2 is considerably longer for the smaller $\Mcore$$_{,1}$. 
 For instance, for the nominal case and $\Sigma_1$= 25 \gcmcuad, Phase 2 for $\Mcore$$_{,1}$= 4 \ME \, lasts 5.3 Myr and for $\Mcore$$_{,1}$=17 \ME, 0.6 Myr. This is the reason that the averaged heavy-element accretion rate is always higher for the larger core mass, as indicated in Figure \ref{maps_nominal}.
 
Table \ref{table_Mz_MMSN} provides the same information as \ref{table_Mz} but for the runs shown in Fig.\ref{MMSN}; where the initial surface density of planetesimals is fixed in terms of the MMSN values \citep{Hayashi81}:
$\Sigma_{1, \rm{MMSN}} = 30.6   f_{\rm ice} \, (a/{\rm au})^{-1.5} $, where $ f_{\rm ice} = 0.5$  inside the water iceline (at $a= 1$ au in our simulations) and 1 for larger semi-major axes.

\begin{table*}
\caption{Heavy elements accreted in each phase of Jupiter's growth (in Earth masses), for each range of initial conditions of ($\Sigma_1$, $\Mcore$$_{,1}$)}              
\label{table_Mz}      
\centering                                      
\begin{tabular}{c c c c c c c  }          
\hline\hline                        
case & $\Sigma_1$ [\gcmcuad]& $\Mcore$$_{,1}$ [\ME]& Phase 2 & Phase 3, attached & Phase 3, detached (OG) &  Phase 3, detached (SI08) \\    
\hline                                   
    		& 5.0 & 11 & 6.8  & 0.38 & 0.50 & 2.4 \\  
     Nominal     &  & 14 & 6.1 & 0.29 & 0.32 & 2.2 \\     
    		& 25 & 4.0 & 23 & 0.56  & 2.1 & 14 \\
		&  & 17 & 15 & 0.43 & 0.94 & 9.7 \\
    \hline    
 			      & 1.0 & 12 & 1.7  &  0.19 & 0.22 & 0.37 \\  
   $\rp$ =10 \rm{km} & 18 & 4.0 & 29 & 0.62  & 2.0 & 5.3\\   
		            &  & 10 & 26 & 0.50 & 1.6 & 5.6\\
     \hline    
                               & 5.0 & 7.0 & 5.9  & 0.40  &0.38 & 2.8 \\  
  Reduced opacity  &  & 9.0 & 5.6  & 0.48  & 0.35 & 2.7 \\     
    		& 25 & 4.0 & 20  & 0.70  & 1.7 & 15 \\
		&  & 15 & 14 & 0.30 & 0.79 & 11 \\	   
     \hline    
            		              & 5.0 & 10 & 5.7  & 0.40  & 0.67 & 2.9 \\  
   Envelope enrichment   &  & 13 & 4.9  & 0.30  & 0.50 &  2.7 \\     
    		                       & 25 & 4.0 & 19 & -  & 2.1 & 18 \\
		                       &  & 17 &  11 & - & 1.6 & 16 \\   
      \hline    	
		 & 5.0   & 13 & 0.37  &  0.07 & 0.53 & 0.056 \\  
   a = 1 au  &      & 18 & 0.37 & 0.05  & 0.35 &  0.074\\     
    		 & 25 & 13 &1.8  & 0.45  & 2.6 & 0.25\\
		 &       & 21 & 2.1 & 0.30 & 1.3 & 0.18\\    
      \hline    
        		& 5.0 & 11 &  7.3 & 0.38 & 0.48 & 11 \\
  a = 10 au    &  & 12 & 6.7  & 0.36 & 2.2 & 11 \\  
 		& 25 & 5.0 & 20 & 0.44 &  9.6 & 50 \\     
		&  & 16 & 13 & 0.52 & 6.6 & 41 \\    
	\hline                                             
\end{tabular}
\end{table*}


\begin{table*}
\caption{Heavy elements accreted in each phase of Jupiter's growth (in Earth masses), for the simulations corresponding to Fig.\ref{MMSN}}              
\label{table_Mz_MMSN}      
\centering                                      
\begin{tabular}{c c c c c c c  }          
\hline\hline                        
case & a [au] &$\Sigma_1$ [\gcmcuad]& $\Mcore$$_{,1}$ [\ME]& Phase 2 & Phase 3, attached & Phase 3, detached (SI08) \\    
\hline                                   
    		&1 &  30.60 & 12 &  2.2 & 0.59 & 0.29  \\  
2 MMSN    &  &   & 16 & 2.4 &  0.49  & 0.27 \\     
    		& 3 & 11.78 & 9.0 & 7.5 & 0.66  & 1.6\\
		&  &   & 17 &7.0 & 0.36 & 0.97\\
		& 5 & 5.470  & 10  & 7.5 & 0.39 & 2.7\\
		&    & &  14 & 6.6 & 0.34 & 2.5  \\
		& 10  & 1.940 &  -  & - & - & -\\
    \hline    
   
      & 1 & 76.50  & 10 & 5.6 &  1.4  & 0.69 \\%
          & & &  17  & 6.3 &  0.99   & 0.58  \\
5 MMSN   & 3  & 29.45 & 4.0 & 19  & 0.68  & 4.0 \\%
                & & &  17  & 15 & 0.52  & 2.7 \\
  		&  5  & 13.69  & 6.0  & 16 & 0.45 &  7.3\\%
      	        & & & 17  &11  & 0.45 & 5.6\\
	       & 10 &  4.840 & 11 & 7.1 & 0.36 & 10\\
   			    & & & 12  &  6.5 & 0.34  & 10 \\
      \hline
  		 & 1 & 153.0  &6.0  & 12 & 2.7  & 1.3 \\%
    		  & & &  17 & 14 & 1.4  & 0.94 \\
 10 MMSN & 3  & 58.30 & 4.0  & 31 & 0.59   &  6.5 \\%
  				& & & 17 &  23 & 0.41 & 6.1 \\
 		  &  5  & 27.40 & 4.0  & 24 & 0.55 & 15 \\%
  			& & & 17 &  16 & 0.46  & 11 \\ 
	        &  10 & 9.680 & 9.0  &  12  & 0.40 & 20 \\
 		&     &     &     13 & 9.2 &  0.38 & 18  \\
	\hline                                             
\end{tabular}

\end{table*}

\end{appendix}

\bibliographystyle{aa}
\bibliography{lit_2019}

\begin{thebibliography}{82}
\expandafter\ifx\csname natexlab\endcsname\relax\def\natexlab#1{#1}\fi

\bibitem[{{Alibert} {et~al.}(2013){Alibert}, {Carron}, {Fortier}, {Pfyffer},
  {Benz}, {Mordasini}, \& {Swoboda}}]{Alibert13}
{Alibert}, Y., {Carron}, F., {Fortier}, A., {et~al.} 2013, \aap, 558, A109

\bibitem[{{Alibert} {et~al.}(2005{\natexlab{a}}){Alibert}, {Mordasini}, {Benz},
  \& {Winisdoerffer}}]{Alibert05}
{Alibert}, Y., {Mordasini}, C., {Benz}, W., \& {Winisdoerffer}, C.
  2005{\natexlab{a}}, \aap, 434, 343

\bibitem[{{Alibert} {et~al.}(2005{\natexlab{b}}){Alibert}, {Mousis}, \&
  {Benz}}]{Alibert05a}
{Alibert}, Y., {Mousis}, O., \& {Benz}, W. 2005{\natexlab{b}}, \apjl, 622, L145

\bibitem[{{Alibert} {et~al.}(2005{\natexlab{c}}){Alibert}, {Mousis},
  {Mordasini}, \& {Benz}}]{Alibert05b}
{Alibert}, Y., {Mousis}, O., {Mordasini}, C., \& {Benz}, W. 2005{\natexlab{c}},
  \apjl, 626, L57

\bibitem[{Alibert {et~al.}(2018)Alibert, Venturini, Helled, Ataiee, Burn,
  Senecal, Benz, Mayer, Mordasini, Quanz, \& Sch{\"o}nb{\"a}chler}]{A18}
Alibert, Y., Venturini, J., Helled, R., {et~al.} 2018, Nature Astronomy

\bibitem[{{Andrews} {et~al.}(2010){Andrews}, {Wilner}, {Hughes}, {Qi}, \&
  {Dullemond}}]{Andrews10}
{Andrews}, S.~M., {Wilner}, D.~J., {Hughes}, A.~M., {Qi}, C., \& {Dullemond},
  C.~P. 2010, \apj, 723, 1241

\bibitem[{{Ataiee} {et~al.}(2018){Ataiee}, {Baruteau}, {Alibert}, \&
  {Benz}}]{Sareh18}
{Ataiee}, S., {Baruteau}, C., {Alibert}, Y., \& {Benz}, W. 2018, \aap, 615,
  A110

\bibitem[{{Batalha} {et~al.}(2013){Batalha}, {Rowe}, {Bryson}, {Barclay},
  {Burke}, {Caldwell}, {Christiansen}, {Mullally}, {Thompson}, {Brown},
  {Dupree}, {Fabrycky}, {Ford}, {Fortney}, {Gilliland}, {Isaacson}, {Latham},
  {Marcy}, {Quinn}, {Ragozzine}, {Shporer}, {Borucki}, {Ciardi}, {Gautier},
  {Haas}, {Jenkins}, {Koch}, {Lissauer}, {Rapin}, {Basri}, {Boss}, {Buchhave},
  {Carter}, {Charbonneau}, {Christensen-Dalsgaard}, {Clarke}, {Cochran},
  {Demory}, {Desert}, {Devore}, {Doyle}, {Esquerdo}, {Everett}, {Fressin},
  {Geary}, {Girouard}, {Gould}, {Hall}, {Holman}, {Howard}, {Howell},
  {Ibrahim}, {Kinemuchi}, {Kjeldsen}, {Klaus}, {Li}, {Lucas}, {Meibom},
  {Morris}, {Pr{\v s}a}, {Quintana}, {Sanderfer}, {Sasselov}, {Seader},
  {Smith}, {Steffen}, {Still}, {Stumpe}, {Tarter}, {Tenenbaum}, {Torres},
  {Twicken}, {Uddin}, {Van Cleve}, {Walkowicz}, \& {Welsh}}]{Batalha13}
{Batalha}, N.~M., {Rowe}, J.~F., {Bryson}, S.~T., {et~al.} 2013, \apjs, 204, 24

\bibitem[{{Bell} \& {Lin}(1994)}]{BL94}
{Bell}, K.~R. \& {Lin}, D.~N.~C. 1994, \apj, 427, 987

\bibitem[{{Birnstiel} {et~al.}(2012){Birnstiel}, {Klahr}, \&
  {Ercolano}}]{Birnstiel12}
{Birnstiel}, T., {Klahr}, H., \& {Ercolano}, B. 2012, \aap, 539, A148

\bibitem[{{Bitsch} {et~al.}(2015){Bitsch}, {Johansen}, {Lambrechts}, \&
  {Morbidelli}}]{Bitsch15a}
{Bitsch}, B., {Johansen}, A., {Lambrechts}, M., \& {Morbidelli}, A. 2015, \aap,
  575, A28

\bibitem[{{Bitsch} {et~al.}(2018){Bitsch}, {Morbidelli}, {Johansen}, {Lega},
  {Lambrechts}, \& {Crida}}]{Bitsch18}
{Bitsch}, B., {Morbidelli}, A., {Johansen}, A., {et~al.} 2018, \aap, 612, A30

\bibitem[{{Bodenheimer} \& {Pollack}(1986{\natexlab{a}})}]{Bodenheimer1986}
{Bodenheimer}, P. \& {Pollack}, J.~B. 1986{\natexlab{a}}, \icarus, 67, 391

\bibitem[{{Bodenheimer} \& {Pollack}(1986{\natexlab{b}})}]{BP86}
{Bodenheimer}, P. \& {Pollack}, J.~B. 1986{\natexlab{b}}, \icarus, 67, 391

\bibitem[{{Bodenheimer} {et~al.}(2018){Bodenheimer}, {Stevenson}, {Lissauer},
  \& {D'Angelo}}]{Bodenheimer2018}
{Bodenheimer}, P., {Stevenson}, D.~J., {Lissauer}, J.~J., \& {D'Angelo}, G.
  2018, \apj, 868, 138

\bibitem[{{Bolton} {et~al.}(2017){Bolton}, {Lunine}, {Stevenson}, {Connerney},
  {Levin}, {Owen}, {Bagenal}, {Gautier}, {Ingersoll}, {Orton}, {Guillot},
  {Hubbard}, {Bloxham}, {Coradini}, {Stephens}, {Mokashi}, {Thorne}, \&
  {Thorpe}}]{Bolton17}
{Bolton}, S.~J., {Lunine}, J., {Stevenson}, D., {et~al.} 2017, \ssr, 213, 5

\bibitem[{{Chambers}(2006)}]{Chambers06}
{Chambers}, J. 2006, \icarus, 180, 496

\bibitem[{{Cumming} {et~al.}(2018){Cumming}, {Helled}, \&
  {Venturini}}]{Cumming18}
{Cumming}, A., {Helled}, R., \& {Venturini}, J. 2018, \mnras, 477, 4817

\bibitem[{{Debras} \& {Chabrier}(2019)}]{Debras19}
{Debras}, F. \& {Chabrier}, G. 2019, \apj, 872, 100

\bibitem[{{Desch} {et~al.}(2018){Desch}, {Kalyaan}, \& {O'D.
  Alexander}}]{Desch18}
{Desch}, S.~J., {Kalyaan}, A., \& {O'D. Alexander}, C.~M. 2018, \apjs, 238, 11

\bibitem[{{Dr{\c a}{\.z}kowska} {et~al.}(2016){Dr{\c a}{\.z}kowska}, {Alibert},
  \& {Moore}}]{Drazkowska16}
{Dr{\c a}{\.z}kowska}, J., {Alibert}, Y., \& {Moore}, B. 2016, \aap, 594, A105

\bibitem[{{Fortier} {et~al.}(2013){Fortier}, {Alibert}, {Carron}, {Benz}, \&
  {Dittkrist}}]{Fortier13}
{Fortier}, A., {Alibert}, Y., {Carron}, F., {Benz}, W., \& {Dittkrist}, K.-M.
  2013, \aap, 549, A44

\bibitem[{{Fortier} {et~al.}(2007){Fortier}, {Benvenuto}, \&
  {Brunini}}]{Fortier07}
{Fortier}, A., {Benvenuto}, O.~G., \& {Brunini}, A. 2007, \aap, 473, 311

\bibitem[{{Fortier} {et~al.}(2009){Fortier}, {Benvenuto}, \&
  {Brunini}}]{Fortier09}
{Fortier}, A., {Benvenuto}, O.~G., \& {Brunini}, A. 2009, \aap, 500, 1249

\bibitem[{{Freedman} {et~al.}(2014){Freedman}, {Lustig-Yaeger}, {Fortney},
  {Lupu}, {Marley}, \& {Lodders}}]{Freedman14}
{Freedman}, R.~S., {Lustig-Yaeger}, J., {Fortney}, J.~J., {et~al.} 2014, \apjs,
  214, 25

\bibitem[{{Guilera} {et~al.}(2010){Guilera}, {Brunini}, \&
  {Benvenuto}}]{Guilera10}
{Guilera}, O.~M., {Brunini}, A., \& {Benvenuto}, O.~G. 2010, \aap, 521, A50

\bibitem[{{Guilera} {et~al.}(2014){Guilera}, {de El{\'{\i}}a}, {Brunini}, \&
  {Santamar{\'{\i}}a}}]{Guilera14}
{Guilera}, O.~M., {de El{\'{\i}}a}, G.~C., {Brunini}, A., \&
  {Santamar{\'{\i}}a}, P.~J. 2014, \aap, 565, A96

\bibitem[{{Guilera} {et~al.}(2011){Guilera}, {Fortier}, {Brunini}, \&
  {Benvenuto}}]{Guilera11}
{Guilera}, O.~M., {Fortier}, A., {Brunini}, A., \& {Benvenuto}, O.~G. 2011,
  \aap, 532, A142

\bibitem[{{Guillot} {et~al.}(2006){Guillot}, {Santos}, {Pont}, {Iro}, {Melo},
  \& {Ribas}}]{Guillot06}
{Guillot}, T., {Santos}, N.~C., {Pont}, F., {et~al.} 2006, \aap, 453, L21

\bibitem[{{Hartmann} {et~al.}(1998){Hartmann}, {Calvet}, {Gullbring}, \&
  {D'Alessio}}]{Hartmann98}
{Hartmann}, L., {Calvet}, N., {Gullbring}, E., \& {D'Alessio}, P. 1998, \apj,
  495, 385

\bibitem[{{Hasegawa} {et~al.}(2018){Hasegawa}, {Bryden}, {Ikoma}, {Vasisht}, \&
  {Swain}}]{Hasegawa18}
{Hasegawa}, Y., {Bryden}, G., {Ikoma}, M., {Vasisht}, G., \& {Swain}, M. 2018,
  ArXiv e-prints

\bibitem[{{Hayashi}(1981)}]{Hayashi81}
{Hayashi}, C. 1981, Progress of Theoretical Physics Supplement, 70, 35

\bibitem[{{Hori} \& {Ikoma}(2011)}]{HI11}
{Hori}, Y. \& {Ikoma}, M. 2011, \mnras, 416, 1419

\bibitem[{{Hubickyj} {et~al.}(2005){Hubickyj}, {Bodenheimer}, \&
  {Lissauer}}]{Hubi05}
{Hubickyj}, O., {Bodenheimer}, P., \& {Lissauer}, J.~J. 2005, \icarus, 179, 415

\bibitem[{{Ida} \& {Lin}(2004)}]{IdaLin04}
{Ida}, S. \& {Lin}, D.~N.~C. 2004, \apj, 604, 388

\bibitem[{{Ikoma} {et~al.}(2000){Ikoma}, {Nakazawa}, \& {Emori}}]{Ikoma00}
{Ikoma}, M., {Nakazawa}, K., \& {Emori}, H. 2000, \apj, 537, 1013

\bibitem[{{Inaba} \& {Ikoma}(2003)}]{InabaIkoma03}
{Inaba}, S. \& {Ikoma}, M. 2003, \aap, 410, 711

\bibitem[{{Inaba} {et~al.}(2001){Inaba}, {Tanaka}, {Nakazawa}, {Wetherill}, \&
  {Kokubo}}]{Inaba01}
{Inaba}, S., {Tanaka}, H., {Nakazawa}, K., {Wetherill}, G.~W., \& {Kokubo}, E.
  2001, \icarus, 149, 235

\bibitem[{{Johansen} {et~al.}(2014){Johansen}, {Blum}, {Tanaka}, {Ormel},
  {Bizzarro}, \& {Rickman}}]{JohansenPPVI}
{Johansen}, A., {Blum}, J., {Tanaka}, H., {et~al.} 2014, Protostars and Planets
  VI, 547

\bibitem[{{Kenyon} \& {Bromley}(2012)}]{Kenyon12}
{Kenyon}, S.~J. \& {Bromley}, B.~C. 2012, \aj, 143, 63

\bibitem[{{Kobayashi} {et~al.}(2010){Kobayashi}, {Tanaka}, {Krivov}, \&
  {Inaba}}]{Kobayashi10}
{Kobayashi}, H., {Tanaka}, H., {Krivov}, A.~V., \& {Inaba}, S. 2010, Icarus,
  209, 836

\bibitem[{{Kokubo} \& {Ida}(1998)}]{KI98}
{Kokubo}, E. \& {Ida}, S. 1998, \icarus, 131, 171

\bibitem[{{Kruijer} {et~al.}(2017){Kruijer}, {Burkhardt}, {Budde}, \&
  {Kleine}}]{Kruijer17}
{Kruijer}, T.~S., {Burkhardt}, C., {Budde}, G., \& {Kleine}, T. 2017,
  Proceedings of the National Academy of Science, 114, 6712

\bibitem[{{Lambrechts} \& {Johansen}(2014)}]{LJ14}
{Lambrechts}, M. \& {Johansen}, A. 2014, \aap, 572, A107

\bibitem[{{Lambrechts} {et~al.}(2014){Lambrechts}, {Johansen}, \&
  {Morbidelli}}]{Lambrechts14}
{Lambrechts}, M., {Johansen}, A., \& {Morbidelli}, A. 2014, \aap, 572, A35

\bibitem[{{Leconte} \& {Chabrier}(2012)}]{Leconte12}
{Leconte}, J. \& {Chabrier}, G. 2012, \aap, 540, A20

\bibitem[{{Lichtenberg} {et~al.}(2019){Lichtenberg}, {Golabek}, {Burn},
  {Meyer}, {Alibert}, {Gerya}, \& {Mordasini}}]{Lichtenberg19}
{Lichtenberg}, T., {Golabek}, G.~J., {Burn}, R., {et~al.} 2019, Nature
  Astronomy

\bibitem[{{Lin} {et~al.}(2018){Lin}, {Lee}, \& {Chiang}}]{Lin18}
{Lin}, J.~W., {Lee}, E.~J., \& {Chiang}, E. 2018, \mnras, 480, 4338

\bibitem[{{Lissauer} {et~al.}(2009){Lissauer}, {Hubickyj}, {D'Angelo}, \&
  {Bodenheimer}}]{Lissauer09}
{Lissauer}, J.~J., {Hubickyj}, O., {D'Angelo}, G., \& {Bodenheimer}, P. 2009,
  \icarus, 199, 338

\bibitem[{{Lozovsky} {et~al.}(2017){Lozovsky}, {Helled}, {Rosenberg}, \&
  {Bodenheimer}}]{Lozovsky17}
{Lozovsky}, M., {Helled}, R., {Rosenberg}, E.~D., \& {Bodenheimer}, P. 2017,
  \apj, 836, 227

\bibitem[{{Masset} \& {Snellgrove}(2001)}]{Masset01}
{Masset}, F. \& {Snellgrove}, M. 2001, \mnras, 320, L55

\bibitem[{{Miller} \& {Fortney}(2011)}]{MF11}
{Miller}, N. \& {Fortney}, J.~J. 2011, \apjl, 736, L29

\bibitem[{{Morbidelli} {et~al.}(2016){Morbidelli}, {Bitsch}, {Crida},
  {Gounelle}, {Guillot}, {Jacobson}, {Johansen}, {Lambrechts}, \&
  {Lega}}]{Morby16}
{Morbidelli}, A., {Bitsch}, B., {Crida}, A., {et~al.} 2016, \icarus, 267, 368

\bibitem[{{Morbidelli} {et~al.}(2009){Morbidelli}, {Bottke}, {Nesvorn{\'y}}, \&
  {Levison}}]{Morby09}
{Morbidelli}, A., {Bottke}, W.~F., {Nesvorn{\'y}}, D., \& {Levison}, H.~F.
  2009, \icarus, 204, 558

\bibitem[{{Mordasini}(2013)}]{Mordasini13}
{Mordasini}, C. 2013, \aap, 558, A113

\bibitem[{{Mordasini} {et~al.}(2014){Mordasini}, {Klahr}, {Alibert}, {Miller},
  \& {Henning}}]{Mord14}
{Mordasini}, C., {Klahr}, H., {Alibert}, Y., {Miller}, N., \& {Henning}, T.
  2014, \aap, 566, A141

\bibitem[{{Mordasini} {et~al.}(2016){Mordasini}, {van Boekel}, {Molli{\`e}re},
  {Henning}, \& {Benneke}}]{Mord16}
{Mordasini}, C., {van Boekel}, R., {Molli{\`e}re}, P., {Henning}, T., \&
  {Benneke}, B. 2016, \apj, 832, 41

\bibitem[{{Movshovitz} {et~al.}(2010){Movshovitz}, {Bodenheimer}, {Podolak}, \&
  {Lissauer}}]{Movshovitz10}
{Movshovitz}, N., {Bodenheimer}, P., {Podolak}, M., \& {Lissauer}, J.~J. 2010,
  \icarus, 209, 616

\bibitem[{{Movshovitz} \& {Podolak}(2008)}]{Movshovitz08}
{Movshovitz}, N. \& {Podolak}, M. 2008, \icarus, 194, 368

\bibitem[{{Oka} {et~al.}(2011){Oka}, {Nakamoto}, \& {Ida}}]{Oka11}
{Oka}, A., {Nakamoto}, T., \& {Ida}, S. 2011, \apj, 738, 141

\bibitem[{Ormel(2017)}]{Ormel17}
Ormel, C.~W. 2017, Proceedings of the Sant Cugat Forum on Astrophysics
  (Springer)

\bibitem[{{Owen} \& {Wu}(2017)}]{Owen17}
{Owen}, J.~E. \& {Wu}, Y. 2017, \apj, 847, 29

\bibitem[{{Pirani} {et~al.}(2019){Pirani}, {Johansen}, {Bitsch}, {Mustill}, \&
  {Turrini}}]{Pirani19}
{Pirani}, S., {Johansen}, A., {Bitsch}, B., {Mustill}, A.~J., \& {Turrini}, D.
  2019, \aap, 623, A169

\bibitem[{{Pollack} {et~al.}(1996){Pollack}, {Hubickyj}, {Bodenheimer},
  {Lissauer}, {Podolak}, \& {Greenzweig}}]{P96}
{Pollack}, J.~B., {Hubickyj}, O., {Bodenheimer}, P., {et~al.} 1996, \icarus,
  124, 62

\bibitem[{{Saumon} {et~al.}(1995){Saumon}, {Chabrier}, \& {van Horn}}]{SCVH}
{Saumon}, D., {Chabrier}, G., \& {van Horn}, H.~M. 1995, \apjs, 99, 713

\bibitem[{Shibata {et~al.}(2019)Shibata, Helled, \& Ikoma}]{ShibataEPSC}
Shibata, S., Helled, R., \& Ikoma, M. 2019, in European Planetary Science
  Congress, EPSC--DPS2019--471--1, 2019

\bibitem[{{Shibata} \& {Ikoma}(2019)}]{Shibata19}
{Shibata}, S. \& {Ikoma}, M. 2019, \mnras, 487, 4510

\bibitem[{{Shiraishi} \& {Ida}(2008)}]{SI08}
{Shiraishi}, M. \& {Ida}, S. 2008, \apj, 684, 1416

\bibitem[{{Simon} {et~al.}(2017){Simon}, {Armitage}, {Youdin}, \&
  {Li}}]{Simon17}
{Simon}, J.~B., {Armitage}, P.~J., {Youdin}, A.~N., \& {Li}, R. 2017, \apjl,
  847, L12

\bibitem[{Suzuki {et~al.}(2018)Suzuki, Bennett, Ida, Mordasini, Bhattacharya,
  Bond, Donachie, Fukui, Hirao, Koshimoto, Miyazaki, Nagakane, Ranc,
  Rattenbury, Sumi, Alibert, \& Lin}]{Suzuki18}
Suzuki, D., Bennett, D.~P., Ida, S., {et~al.} 2018, The Astrophysical Journal,
  869, L34

\bibitem[{{Tanigawa} \& {Watanabe}(2002)}]{Tanigawa02}
{Tanigawa}, T. \& {Watanabe}, S.-i. 2002, \apj, 580, 506

\bibitem[{{Thiabaud} {et~al.}(2015){Thiabaud}, {Marboeuf}, {Alibert}, {Leya},
  \& {Mezger}}]{Amaury15}
{Thiabaud}, A., {Marboeuf}, U., {Alibert}, Y., {Leya}, I., \& {Mezger}, K.
  2015, \aap, 574, A138

\bibitem[{{Thommes} {et~al.}(2003){Thommes}, {Duncan}, \&
  {Levison}}]{Thommes03}
{Thommes}, E.~W., {Duncan}, M.~J., \& {Levison}, H.~F. 2003, \icarus, 161, 431

\bibitem[{{Thorngren} {et~al.}(2016){Thorngren}, {Fortney}, {Murray-Clay}, \&
  {Lopez}}]{T16}
{Thorngren}, D.~P., {Fortney}, J.~J., {Murray-Clay}, R.~A., \& {Lopez}, E.~D.
  2016, \apj, 831, 64

\bibitem[{{Valletta} \& {Helled}(2019)}]{Valletta19}
{Valletta}, C. \& {Helled}, R. 2019, \apj, 871, 127

\bibitem[{{Vazan} {et~al.}(2018){Vazan}, {Helled}, \& {Guillot}}]{Vazan18}
{Vazan}, A., {Helled}, R., \& {Guillot}, T. 2018, \aap, 610, L14

\bibitem[{{Venturini} {et~al.}(2016){Venturini}, {Alibert}, \&
  {Benz}}]{Venturini16}
{Venturini}, J., {Alibert}, Y., \& {Benz}, W. 2016, \aap, 596, A90

\bibitem[{{Venturini} \& {Helled}(2017)}]{Venturini17}
{Venturini}, J. \& {Helled}, R. 2017, \apj, 848, 95

\bibitem[{{Wahl} {et~al.}(2017){Wahl}, {Hubbard}, {Militzer}, {Guillot},
  {Miguel}, {Movshovitz}, {Kaspi}, {Helled}, {Reese}, {Galanti}, {Levin},
  {Connerney}, \& {Bolton}}]{Wahl17}
{Wahl}, S.~M., {Hubbard}, W.~B., {Militzer}, B., {et~al.} 2017, \grl, 44, 4649

\bibitem[{{Walsh} {et~al.}(2011){Walsh}, {Morbidelli}, {Raymond}, {O'Brien}, \&
  {Mandell}}]{Walsh11}
{Walsh}, K.~J., {Morbidelli}, A., {Raymond}, S.~N., {O'Brien}, D.~P., \&
  {Mandell}, A.~M. 2011, \nat, 475, 206

\bibitem[{{Weidenschilling}(2011)}]{Weiden11}
{Weidenschilling}, S.~J. 2011, \icarus, 214, 671

\bibitem[{{Zhou} \& {Lin}(2007)}]{Zhou2007}
{Zhou}, J.-L. \& {Lin}, D. N.~C. 2007, \apj, 666, 447

\end{thebibliography}

\end{document}